\documentclass[12pt,preprint]{aastex}

\usepackage{graphicx}
\providecommand{\tabularnewline}{\\}

\begin{document}




\title{Dynamics of Rotating Accretion Flows Irradiated by a Quasar}

\author{Daniel Proga\altaffilmark{1}, Jeremiah P. Ostriker\altaffilmark{2}, 
and Ryuichi Kurosawa\altaffilmark{1}}
\altaffiltext{1}{Department of Physics, University of Nevada, Las Vegas,
NV 89154, USA: dproga@physics.unlv.edu, rk@physics.unlv.edu}
\altaffiltext{2}{Princeton University Observatory, Princeton, USA:
ostriker@astro.princeton.edu}
\def\LSUN{\rm L_{\odot}}
\def\MSUN{\rm M_{\odot}}
\def\RSUN{\rm R_{\odot}} 
\def\MSUNYR{\rm M_{\odot}\,yr^{-1}}
\def\MSUNS{\rm M_{\odot}\,s^{-1}}
\def\MDOT{\dot{M}}

\newbox\grsign \setbox\grsign=\hbox{$>$} \newdimen\grdimen \grdimen=\ht\grsign
\newbox\simlessbox \newbox\simgreatbox
\setbox\simgreatbox=\hbox{\raise.5ex\hbox{$>$}\llap
     {\lower.5ex\hbox{$\sim$}}}\ht1=\grdimen\dp1=0pt
\setbox\simlessbox=\hbox{\raise.5ex\hbox{$<$}\llap
     {\lower.5ex\hbox{$\sim$}}}\ht2=\grdimen\dp2=0pt
\def\simgreat{\mathrel{\copy\simgreatbox}}
\def\simless{\mathrel{\copy\simlessbox}}

\begin{abstract}
We study the axisymmetric, time-dependent hydrodynamics of rotating flows 
that are under the influence of supermassive black hole gravity and 
radiation from  an accretion disk surrounding the black hole. This work is an extension of the earlier work 
presented  by Proga, where  nonrotating flows were studied. Here, we consider 
effects of rotation, a position-dependent radiation temperature, 
density at large radii, and uniform X-ray background radiation.
As in the non-rotating case,  the rotating flow settles 
into a configuration with two components
(1) an equatorial inflow and
(2) a bipolar inflow/outflow with the outflow leaving the system
along the pole. However, with rotation the flow does not always reach 
a steady state. In addition, 
rotation reduces the outflow collimation and 
the outward flux of mass and kinetic energy. Moreover rotation 
increases  the outward flux of the thermal energy and
can lead to fragmentation and time-variability of the outflow.
We also show that a position-dependent radiation temperature
can significantly change the flow solution. In particular,
the inflow in the equatorial region can be replaced 
by a thermally driven outflow.
Generally, as it have been discussed and shown in the past, we find 
that self-consistently determined preheating/cooling from  
the quasar radiation can significantly reduce the rate at which 
the central BH is fed with matter.
However, our results emphasize also a little 
appreciated feature.  Namely, quasar radiation drives
a non-spherical, multi-temperature and very dynamic flow.
These effects become dominant for luminosities in excess of
0.01 of the Eddington luminosity.
\end{abstract}


\section{Introduction}

The key property of Active Galactic Nuclei (AGN) is that they emit
enormous amount of electromagnetic radiation over a very broad
energy range.
The AGN central location in 
their host galaxies imply that AGN radiation can play a very important
role in determining the ionization structure and dynamics
of matter not only near the AGN but also on larger,
galactic and even intergalactic scales
(Ciotti \& Ostriker, 1997, 2001, 2007; 
Silk \& Rees 1998;
King 2003; 
Murray et al. 2005;
Sazonov et al. 2005; Springel et at. 2005;
Hopkins et al. 2005; Wang et al. 2006; 
Fabian et al. 2006;
Thacker et al. 2006,
and references therein). 

In the first paper of this series (Proga 2007, Paper~I hereafter), we reported 
on results from our first phase of gas dynamics studies in AGN on sub- and 
parsec-scales. This is a complex problem as it involves many aspects of physics
such as multi-dimensional fluid dynamics, radiative and magnetic processes. 
Therefore, our approach was to set up simulations as simple as possible and 
to start with exploring  the effects of the X-ray heating 
[important in the so-called preheated accretion, 
(e.g., Ostriker et al. 1976; Park \& Ostriker 2001, 2007)]
and radiation 
pressure on gas that is gravitationally captured by a black hole (BH). 
We adopted the numerical methods developed by Proga et al. (2000, 
PSK00 hereafter) for studying radiation driven disk winds in AGN. 
Generally, our simulations cover a relatively unexplored range of the distance
from the central BH: we end where models of galaxies begin 
(e.g., Ciotti \& Ostriker 2007; Springel et at. 2005) and we begin
where models of BH accretion end (e.g., Hawley \& Balbus 2002; Ohsuga 2007).

In Paper~I, we presented results from axisymmetric time-dependent 
hydrodynamical (HD) calculations of gas flows. The flows were non-rotating and 
exposed to quasar radiation. We took into account X-ray heating and 
the radiation force due to electron scattering and spectral lines. To compute 
the radiation field, we considered an optically thick, geometrically thin, 
standard accretion disk as a source of UV photons and 
a spherical central object
as a source of X-rays (a corona). 
The gas temperature, $T$ and ionization state in the flow 
were calculated self-consistently from the photoionization 
and heating rate of the central object.

We found that, for a $10^8~\MSUN$ black hole with an accretion luminosity
of 0.6 of the Eddington luminosity, the flow settles 
into a steady state and has two components:
(1) an equatorial inflow and
(2) a bipolar inflow/outflow with the outflow leaving the system
along the disk rotational axis. The inflow is a realization
of a Bondi-like accretion flow. The second component
is an example of a non-radial accretion flow which 
becomes an outflow once it is pushed close to the rotational axis 
where the radiation pressure accelerate it outward. In some cases
the outflow is heated by radiation so that it can be accelerated
also by thermal expansion.
Our main result was that the existence of
the above two flow components
is robust to the outer boundary conditions and
the geometry and spectral energy distribution of the radiation field.
However, the flow properties are not robust.
In particular, the outflow power and collimation
is higher for the radiation dominated by the UV/disk emission
than for the radiation dominated by the X-ray/central engine emission.
Our most intriguing result was that a very narrow outflow driven by radiation 
pressure on lines can carry more energy and mass than 
a broad outflow driven by thermal expansion.

Here, we report on results from simulations
that are basically reruns of those presented in Paper~I, but
with inclusion of gas rotation. Our goal is  to assess how rotation changes
the flow solution. In particular, we study how rotation changes
the flow pattern, mass and energy fluxes, and temporal behavior.
We also present results from a new set of simulations 
that illustrate how complex and dynamic the flow evolution 
can be even for relatively simple initial and boundary conditions.
We describe our calculations in Section 2. We present our results in Section 3.
The paper ends, in Section 4, with discussion and our conclusions.

\section{Method}

In this paper we extend the work presented in Paper~I by
relaxing some assumptions and simplifications. 
Our numerical HD calculations
are in most respects as described by Paper~I. Here we only describe 
the key elements of 
the calculations and list the changes we made. 
We refer a reader to Paper~I and PSK00 for details.

We consider an axisymmetric HD flow accreting onto a supermassive BH. 
The flow is non-spherical because it is irradiated 
by an accretion disk. The disk radiation flux, ${\cal F}_{\rm disk}$
is highest along the  disk rotational axis and is gradually decreasing 
with increasing polar angle, $\theta$:  
${\cal F}_{\rm disk}\propto|\cos{\theta}|$.
The flow is also irradiated by a corona.
We  account for some effects of photoionization. In particular, 
we calculate the gas temperature assuming that the gas is optically thin 
to its own cooling radiation. We include the following radiative processes:
Compton heating/cooling, X-ray photoionization heating, and recombination,
bremsstrahlung and line cooling.
We also take into account some  effects 
of photoionization on radiation pressure due to lines (line force). Namely, 
we compute the line force using a value of the photoionization
parameter, $\xi$ and the analytical formulae due to Stevens \& Kallman
(1991). This procedure is computationally efficient and
gives good estimates for the number and opacity distribution of
spectral lines for a given $\xi$ without detail information about
the ionization state (see Stevens \& Kallman 1991).
Additionally, we take into account 
the attenuation of the X-ray radiation by computing the X-ray optical depth 
in the radial direction.  
To be consistent with our gas heating rates where we include 
X-ray photoionization but not UV photoionization,
we do not account for attenuation of the UV radiation.

We assume that the total accretion luminosity, 
$L$ has two components: $L_{\rm disk}=f_{\rm disk} L$ due to the accretion disk
and $L_\ast=f_{\rm \ast} L$ due to the corona. 
For simplicity, we assume that the disk emits only UV photons, whereas
the corona emits only  X-rays i.e.,
the system UV luminosity, $L_{\rm UV}=f_{\rm UV} L=L_{\rm disk}$ 
and
the system X-ray luminosity, $L_{\rm X}=f_{\rm X} L=L_\ast$
(in other words $f_{\rm UV}=f_{\rm disk}$ and $f_{\rm X}=f_\ast$).

With the above simplifications, only the corona
radiation is responsible for ionizing the flow to a very high ionization 
state. In our calculations, 
the corona contributes to the radiation force 
due to electron scattering but does not contribute to line driving
{We note that metal lines in the soft X-ray band may have an appreciable
contribution to the total radiation force in some cases.}
On the other hand, the disk radiation contributes
to the radiation force due to both electron and line scattering.

We perform our calculations in spherical polar coordinates
$(r,\theta,\phi)$ assuming axial symmetry about the rotational axis
of the accretion disk ($\theta=0^o$).

Our computational domain is defined to occupy 
the angular range $0^o \leq \theta \leq 90^o$ and the radial range
$r_{\rm i}~=~500~r_\ast \leq r \leq \ r_{\rm o}~=~ 2.5~\times~10^5~r_\ast$, 
where $r_\ast=3 r_{\rm S}$ is the inner radius of the disk
around a Schwarzschild BH with a mass, $M_{\rm BH}$ and radius 
$r_{\rm S}=2GM_{\rm BH}/c^2$.
The $r-\theta$ domain is discretized into zones.
Our numerical resolution in the $r$ direction consists of 140 zones.
We fix the zone size ratio, 
$dr_{k+1}/dr_{k}=1.04$ (i.e., the zone spacing is increasing outward). 
Gridding in this manner ensures good spatial resolution close to
the inner boundary, $r_{\rm i}$.
In the $\theta$ direction, our numerical resolution consists
of 50 zones.
The zone size ratio is always $d\theta_{l}/d\theta_{l+1} =1.0$ (i.e.,
grid points are equally spaced).

For the initial condition, in Paper~I, we assumed spherical symmetry
and set all HD variables to constant values everywhere in 
the computational domain. Here we will allow the gas to rotate.

\subsection{Gas Rotation}

In simulations with rotation, we break spherical symmetry of the initial and
boundary conditions
by introducing a small, latitude-dependent angular momentum. Namely, for
large radii
we assume the specific angular momentum, $l$,  depends on the polar angle, 
$\theta$, as 
\begin{equation}
l(\theta)=l_0 f(\theta),
\end{equation}
where  $f=1$ on the equator ($\theta=90^\circ$) and monotonically
decreases to zero at the poles (at $\theta=0^\circ$ and $180^\circ$).
The initial distribution of the rotational velocity is:
\begin{equation}
v_\phi(r, \theta)=\left\{ \begin{array}{ll}
                 0    & {\rm for}~~\,~~r~~<~~10^{5}~{r_\ast} ~~\,~~\\
                 l/\sin{\theta}~r    & {\rm for}~~\,~~r~~\ge~10^{5}~{r_\ast .} ~~\,~~\\
\end{array}
\right.
\end{equation}

We express the specific angular momentum on the equator  as
\begin{equation}
l_0=\sqrt{\rm r'_{\rm c}/6}~c~r_\ast ,
\end{equation}
where $r'_{\rm c}$ is the ``circularization radius'' on the equator in units 
of $r_\ast$ for the Newtonian potential 
(i.e.,  $ GM/r^2= v^2_\phi/r$ at $r= r'_{\rm c} r_\ast$).
We adopt two forms for the function $f(\theta)$: 
\begin{equation}
f_1(\theta)=\left\{ \begin{array}{ll}
                 0    & {\rm
                  for}~~\,~~\theta~~<~~\theta_o~~{\rm and}~~\theta >
                 180^\circ -\theta_o  ~~\,~~\\
                 l_0  & {\rm for}~~\,~~\theta_o\le~~\theta~~\le 
                   180^\circ-\theta_o,
\end{array}
\right.
\end{equation}
and
\begin{equation}
f_2(\theta)=1-|\cos\theta|.
\end{equation}

We also follow Paper~I in setting the boundary conditions 
at the outer radius, $r_{\rm o}$:
to represent steady conditions at the outer radial boundary. 
During the evolution of each model, we apply the constraints that 
in the last zone in the radial direction, all HD quantities, expect
the radial velocity, are set to their initial values at all times.
The radial velocity is allowed to float. In this paper
we set $v_\phi(\rm r_o, \theta)$ using eq. 2 whereas in Paper~I,
$v_\phi(\rm r_o, \theta)$ was set to zero consistently with the initial 
conditions.

\subsection{Mean Radiation Energy}

One of the simplifications made in Paper~I was an assumption 
that the radiation temperature, 
$T_{\rm R}$ [or more generally the spectral energy distribution (SED)] 
does not change with the position in the flow and it was set to the temperature
of the isotropic corona radiation, $T_{\rm X}$.
This assumption finds justification in X-ray observations which show that 
quasar radiation heats a low-density gas nearly uniformally, 
on scales comparable to the Bondi radius,
up to an equilibrium Compton temperature of 
about $ 2\times10^7$~K (Sazonov et al. 2005; Allen et al. 2006
and references therein). However, within our theoretical framework,
the SED and $T_{\rm R}$ can change with position. For example,
we assume that the UV/disk radiation decreases with increasing $\theta$
(see eq. 6 in Paper~I). Thus even if
the X-ray/central object radiation does not change 
with $\theta$  (see eq. 7 in Paper~I)
the ratio between the X-ray and
UV fluxes increases with increasing $\theta$. Consequently, our model
predicts that radiation is softer near the poles than near the equator.
Moreover, we take into account the attenuation of the X-ray
radiation but not of the UV radiation. Thus, for a given
$\theta$ the ratio between the X-ray and
UV fluxes can decrease if the X-ray optical depth is high.
To test how important  these effects are, we allow $T_{\rm R}$
to vary with position.

We introduce position-dependence of $T_{\rm R}$ in the following way.
We start with the standard expression for the net Compton heating rate:
\begin{equation}
{\cal L}=n_e \frac{\sigma {\cal F}}{m_e c^2}(k T_{\rm R} -4 k T),
\end{equation}
where $\cal F$ is the radiation flux, $n_e$ is the electron number 
density (other symbols have their usual meaning).

As metioned above, we consider two sources of radiation: a disk that 
emits UV photons with
energy between 0 and 50~eV, and a corona that emits photons with
energy above 50~eV. 
Each of this sources has its own 
mean photon energy, $<\epsilon>$ or equivalently radiation temperature, 
($<\epsilon>=k T_{\rm R}$). The radiation temperature of these two sources
can be computed from:
\begin{equation}
k T_{\rm disk}= \int_{0~eV}^{50~eV} h\nu {\cal F}_{{\rm disk},\nu} d\nu/
\int_{0~eV}^{50~eV} {\cal F}_{{\rm disk},\nu} d\nu
\end{equation}
and
\begin{equation}
k T_{\rm X}=\int_{50~eV}^{\infty} h\nu {\cal F}_{{\rm X},\nu} d\nu/
\int_{50~eV}^{\infty} {\cal F}_{{\rm X},\nu} d\nu.
\end{equation}
We consider these temperatures as additional free parameters.
We  chose the radiation temperature of the disk to be
$2\times 10^4$~K and of the corona to be $2.9\times 10^8$~K.

Having set the radiation temperature of the disk and corona radiation
we can compute the radiation temperature of the total 
radiation field from:
\begin{equation}
T_{\rm R}(\theta)=T_{\rm UV} 
\frac{f_{\rm X} R_{\rm T} \exp(-\tau_{\rm X})+ 
2 f_{\rm UV}  \cos{\theta}}
{f_{\rm X}\exp(-\tau_{\rm X}) + 2 f_{\rm UV}  \cos{\theta}},
\end{equation}
where $R_{\rm T}=T_{\rm X, max}/T_{\rm UV}$.
To obtain the expression above we used the definition of $T_{\rm R}$ and
the formulae for the radiation fluxes from the disk and corona 
(eqs. 6 and 15 in Paper~I).  Practically, to account for 
the position-dependent radiation temperature, we adopt the same
formulae for the radiative heating/cooling rates  as in Paper~I but
replace $T_{\rm X}$ with $T_{R}$ (see eqs. 19 and 20 in PSK).
The seventh column in Table~1 shows the values of the adopted $T_{\rm R}$
or its range for cases when we use eq. 9.

We finish this section with a note about the gas temperature
at  the outer radius.
This temperature is typically set to  
the Compton temperature assuming that the central X-rays heat a gas
up to an equilibrium Compton temperature. 
However, in some  of our simulations
the optical depth toward the radiation source or the local density in the flow
is so high that radiation cannot heat a gas at large radii
to the Compton temperature.
In such cases, there is a mistmatch between
the gas temperature assumed at $r_{\rm o}$ (i.e., $T_{\rm 0}$) 
and that computed for $r$ close to $r_{\rm o}$.

It is possible that gas at large radii is heated not only
by the central source but also by shocks or other sources such as
supernovae. To mimic such sources of heating, we introduce
an uniform background X-ray radiation, ${\cal F}_{\rm X, b}$ in some of our
simulations. We illustrate effects of this radiation by showing results
of one model (run Crbgd), where we assumed
${\cal F}_{\rm X, b}=1.2 \times 10^{7}$~erg~cm$^{-2}$~s$^{-1}$, for which
the gas with relatively low
density is comptonized, e.g., for $\rho=1 \times 10^{-20}$~g~cm$^{-3}$,
$\xi=2.5\times 10^{4}$ (we assume that
$T_{\rm R}$ of the background radiation is the same as of 
the central source, that it $8\times 10^7$~K.)

\section{Results}

As in Paper~I, we assume the mass of the nonrotating BH,
$M_{\rm BH}~=~10^8~\rm \MSUN$ and
the disk inner radius, $r_\ast=~3~r_{\rm S}~=~8.8~\times~10^{13}$~cm throughout
this paper. We compute the total accretion luminosity
as 
$L=\eta \MDOT_{\rm a} c^2=2 \eta G M_{\rm BH} \MDOT_{\rm a}/r_{\rm S}$, 
where $\eta$ and $\MDOT_{\rm a}$ is the rest mass conversion efficiency  
and the mass accretion on BH, respectively. We assume a relatively
high conversion efficiency appropriate for disk accretion  onto
a non-rotating BH, i.e., $\eta=~0.0833$.

We express the accretion luminosity in
in units of the
Eddington luminosity for the Schwarzschild BH, i.e.,
$L_{\rm Edd}= 4 \pi c G M_{\rm BH}/\sigma_e$. We refer to this
normalized luminosity as the Eddington number,
$\Gamma\equiv L/L_{\rm Edd}= (\sigma_e \MDOT_{\rm a})/(8\pi c r_{\rm S}$).

Table~1 summarizes the properties of models from Paper~I 
(runs A, B, B1, B2, B3, and C) and our new 
models (the other models listed in the table). 
Columns (2) to (11) give the input parameters that we varied: 
the Eddington number, $\Gamma$,
the disk contribution to the total luminosity, $f_{\rm disk}$,
the corona contribution to the total luminosity, $f_{\ast}$,
the UV contribution to the total luminosity, $f_{\rm UV}$,
the X-ray contribution to the total luminosity, $f_{\rm X}$,
the radiation temperature, $T_{\rm R}$,
the X-ray background radiation flux, ${\cal F}_{\rm X,b}$,
the gas temperature at the outer boundary, $T_{\rm 0}$,
the gas density at the outer boundary, $\rho_{\rm 0}$,
and the circularization radius, $r'_{\rm c}$.
Columns (12) to (17) give
some of the gross properties of the solutions: 
the mass inflow rate through the outer boundary $\MDOT_{\rm in}(r_{\rm o})$, 
the net mass flux rate through the inner boundary $\MDOT_{\rm net}(r_{\rm i})$, the mass outflow rate through the outer boundary $\MDOT_{\rm out}(r_{\rm o})$, 
the maximum outflow velocity at the outer boundary, $v_{\rm r}$,
the outflow power carried out through the outer boundary
in the form of kinetic energy, $P_{\rm k}(r_{\rm o})$, and
in the form of thermal energy, $P_{\rm th}(r_{\rm o})$.
Table~1 also  explains  our convention of  labeling our runs.
All other model parameters not listed in Table~1 are as in Paper~I.

\subsection{Effects of Gas Rotation}

Simulations without rotation presented in Paper~I, show that
an infalling gas collimates an outflowing gas and that
the collimation increases with increasing radius.
In addition, for a given $\Gamma$, the collimation increases
with decreasing ratio between $f_{\rm X}$ and $f_{\rm UV}$.
Out of three cases explored in Paper~I, case C
with the smallest $f_{\rm UV}/f_{\rm X}$, 
shows the strongest collimation and  
the highest efficiency of turning
an inflow into an outflow.
Fig.~1 in Paper~I and Fig.~1 here show that 
in run~C the gas is siphoned off within a very narrow
channel along the pole. 

In Paper~I, we argued that the collimation, outflow power and other
results  will likely change if one would allow 
for significant gas rotation. One would expect that
the gas will converge toward the equator due to the combination of 
the centrifugal and gravitational forces. This, in turn, will likely 
broaden and weaken 
the outflow in the polar region because less gas will be pushed toward 
the polar region. 

To test effects of rotation, we rerun models presented in Paper~I 
with rotation. We set $r'_{\rm c}=300$ that is near the maximum value
for which the flow will not circularize inside our computational domain.
We do not consider higher $r'_c$ at the moment because we want to avoid
complexities that will result from formation a rotationally
supported torus or disk inside the computational domain 
(e.g., 
Hawley, Smarr \& Wilson 1984a; 1984b; 
Clarke, Karpik \& Henriksen 1985; 
Hawley 1986; 
Molteni et al. 1994;
Ryu et al. 1995; 
Chen et al.  1997;
Toropin et al. 1999; 
Kryukov et al. 2000;
Igumenshchev \& Narayan 2002;
Proga \& Begelamn 2003;
Chakrabarti et al. 2004, and references therein). 
In these exploratory simulations, our choice of high $r'_{\rm c}$,
yielding low $l$, allows us then to study first, relatively simple flows 
and set a stage for modeling more complex flows
with high $l$. We assume
that the circularized gas will accrete onto SMBH on a viscous time scale.
However, as we do not model this part of the flow, we do not consider 
any details of an actual process/es leading
to transport of angular momentum. The transport is 
most likely due to magnetorotational instability (Balbus \& Hawley 1991) 
but a contribution from the photon viscosity can be important in the case 
of high radiation luminosity.

We note that our choice of low $l$ can be relevant to real QSOs because
feeding a SMBH in QSO with gas of high $l$ can be problematic. Namely,
for high $l$, an accretion disk would form at large radii and 
be self-gravitating (e.g., Paczynski 1978; Shlosman \& Begelman 1987). 
Converting
disk material to stars could then starve the SMBH and QSO would be quenched
(e.g., Goodman 2003). After reviewing variety of possibilities, Goodman (2003)
suggested  that QSO disks do not extend beyond a thousand $r_S$
so that they could be gravitationally stable. If so
such the disks must be replenished with gas of small $l$ 
as that we explore here.

Figure~1 presents the results for runs C, CR, and Cr.
For run~CR,
a step function describes  the angular distribution of angular momentum on 
the outer boundary (we set $\theta_0$ to $45^\circ$ in see eq. 4).
For  $45^\circ \le \theta$, we assume that 
the specific angular momentum at 
the outer boundary equals  $l_0$, 
whereas for $\theta < 45^\circ$, $l=0$.
For run Cr, a smooth  function describes  $l$ (see eq. 5).
The figure shows the instantaneous density and temperature
distributions, and the poloidal velocity field of the models.
In addition, it shows the Compton radius corrected for the effects
of radiation pressure due to electrons (see eq. [19] in Paper~I) and
the contours of the Mach number equal to one, 
($M~\equiv~\sqrt{v^2_r+v^2_\theta}/c_s=1$,
where $c_s=\gamma P/\rho$ is the sound speed).

The detailed calculations confirm our general expectations:
compared with the non-rotating case, 
the outflow is less collimated and weaker in the rotating case. 
As Table~1 shows, in the runs with rotation (runs CR and Cr), 
the outflow power, 
$\MDOT_{\rm out}(r_{\rm o})$, and $v_{\rm r}$ are lower than those 
in the run
without rotation (run C; compare also fig. 4 in Paper~1
with Figs.~2 and 3 here).
Another difference is that in runs~CR and Cr,
gas does not cool as much as in run C,
especially at small radii (i.e., $r'<1\times10^5$).
Comparison between runs C, CR, and Cr shows that  
in run CR the solution   is an intermediate one
between runs C and Cr which is not too surprising, given
the step function is an intermediate distribution between zero-$l$
and the one described by function $f_1$. In particular,
the outflow in run CR is less collimated than
in run C and more collimated than in run Cr.
One of new unexpected features that we found in run Cr 
is that the relatively cold 
outflow is fragmented and time-variable.

Although the flow in run Cr settles down into a time-averaged steady state,
it is not as steady as in run C.  An indication of this behavior
can be found in Fig.~3 that shows
three radial  mass flow rates as a function of radius: the net rate,
$\MDOT_{\rm net}(r)$, the inflow rate
$\MDOT_{\rm in}(r)$, and the outflow
rate $\MDOT_{\rm out}(r)$ 
(see eqs. 22, 23, 24 in Paper~I for formal definitions).
For a perfect steady state, one expects
$\MDOT_{\rm net}(r)=\MDOT_{\rm in}(r)+\MDOT_{\rm out}(r)=$const at all radii
as in run C (see Fig. 4 in Paper~I). 
However, in case Cr the above
equation holds only at small radii, $r' \simless 10^4$~K.
We note that contrary to $\MDOT_{\rm out}$,
$\MDOT_{\rm in}$ is a smooth function of radius. Thus the unsteadiness
of in the flow appears to be caused by the unsteady behavior of the
outflow, especially the outflow at large radii where it can
cool down.

We relate  the fragmentation and time-variability of the outflow 
to line force turning on abruptly
when $T$ decreases  below $\sim 5\times 10^4$~K
and turning off when $T$ increases again.
Fig.~1 shows  that $T$ decreases in the regions
where the inflow sharply turns into the outflow. The density there
increases and the gas radiatively cools. The turning on
of line force leads to an enhanced acceleration
of the outflow, but this alone is not sufficient to fragment the outflow.
For example, in runs~C and CR the cold outflow is not fragmented and is
quite steady. Thus there must be another factor/s that may contribute
to fragmentation and time-variability. 
We note that in run~C, the outflow is nearly radial hence
its inner parts shield the outer parts form the central radiation. 
Consequently,
the outflow can not be heated downstream by 
the central radiation. However, in run Cr the outflow is 
not radial and the flow can be heated up downstream
because, as its density decreases during acceleration, it is irradiated
by stronger unattenuated X-ray flux.
There the outflow orientation with respect to the radiation flux 
appears to be one of the key factors causing
fragmentation and time-variability of the outflow.
This conclusion is supported by the fact that even
in run Cr, the outflow is not fragmented at large radii
where the outflow becomes almost radial,  and 
clumps merge with each other.

To show the variable solution in more details, in Fig.~4 
we present a sequence of density maps of
the inner part of the flow in run~Cr 
at four different times. The left panel shows the flow at a time when a clump
brakes from a high density filament at $z' \approx 1.5\times10^4~r_\ast$.
Subsequent panels show how this and other clumps move outward and
how they are stretched. Fig. 4 shows also
formation a new clump (the second panel from the right).
We find that clumps form usually at the same location
(i.e., at $r'\approx 5\times 10^3~r_\ast$ and 
$z'\approx 1.5\times10^4~r_\ast$)
every $10^{11}$~s or so which is of order of a dynamical time scale
at radius where the clumps form. Generally, despite the time-variability,
the instantaneous maps shown in Fig.~1
are representative of run Cr because they
show an example of a large scale inflow and outflow with
continuous production   of small scale clumps that
merge at large radii (i.e., beyond $r' \simgreat 10^5$).

Figs. 3 and 4 show that in runs CR and Cr the outflow power is dominated
by the kinetic energy not the thermal energy. However, comparing
with run~C (see Fig.~4 in Paper~I)
the dominance is not as strong. 

We conclude that in case C, 
rotation reduces the outflow collimation and 
the outward flux of mass and kinetic energy. Rotation also
leads to fragmentation and time-variability of the outflow and an increase
of the outward flux of the thermal energy. As expected, rotation
does not change much the mass inflow rate through the outer
boundary. 

Figs.~5 and 6 show results for case B with and without
rotation, i.e., runs B and Br (see also Fig.~3 in Paper~I).
In this case,
rotational effects are almost the same 
as in case C. The main difference is that in run Br, an outflow
does fragment and the overall flows settles down into a steady state.
This however is understandable: in run Br, radiative
heating is strong and the gas does not cool therefore
line force does not turn on.

\subsection{Effects of the position-dependent radiation temperature}

We return now to case C and consider effects of the position-depenedent
$T_{\rm R}$. Left panels of Fig.~7 and Fig.~8 show results for run Cx.
In comparison with run~C, the outflow in run Cx is broader.
The mass outflow rate in run Cx is only slightly higher than in run~C.
However, the outflow rate almost cancels out the inflow rate
so that the net rate is two orders of magnitude smaller
than the mass flux through the outer boundary. Run Cx is a good example
where AGN irradiation can significantly reduce the rate at which 
the central engine is fed with matter.

In run~Cx, $T_{\rm R}$ is lower near the poles than near the equator.
In addition, $T_{\rm R}$ near the pole is lower in run Cx
than $T_{\rm R}$ in run C. The latter difference
explains why the outflow in run Cx is so strong: the relatively low $T_{\rm R}$
in the polar region leads to a lower gas temperature in this region.
This in turn leads to more mass being pushed towards the pole.
This mass can then be effectively turned into an outflow
because in the polar region the radiation flux is sub-Eddington
and additionally, line-force turns on there because the gas temperature is 
low enough.
In other words, the siphon effect seen in many of our simulations
is very strong in run Cx.

Right panels of Fig.~7 and Fig.~9 show results 
for run Crx that is a rerun of run Cx with rotation.
Comparing these two run, we find that
effects of rotation in the simulations
with the position-dependent $T_{\rm R}$ are similar to
those in the simulations with constant $T_{\rm X}$.
Namely, rotation decreases the degree of the outflow collimation and 
decreases the outward flux of mass and kinetic energy. 
In addition, rotation 
leads to
an increase of the outward flux of the thermal energy. 
In run Crx, the cold outflow is nearly radial and
does not fragment as much as in run Cr.

In this paper, we do not present results for case A with rotation
because in this case the flow is dominated by thermal effects
and rotation does not change much the solution.
However, we present here results for case A with the
position-dependent $T_{\rm R}$ (run Ax) because they show new effects.

Fig.~10 compares results for runs A and Ax. In run Ax, the flow
pattern is different that that seen in runs presented in Paper~I or runs
shown here so far. The dramatic difference is that in run Ax
the equatorial inflow is replaced by an equatorial outflow.
Generally, in run Ax there are  equatorial and polar outflows
both being fed by an inflow of gas at intermediate polar angles.
The equatorial outflow is a simple consequence of the higher
radiation temperature near the equator that leads
to a high gas temperature and an enhanced thermal expansion.
In the polar region, where the gas temperature is lower, an outflow is 
driven by thermal expansion assisted by radiation pressure on electrons.

Overall the flow pattern in run Ax is dominated by the outflow.
This leads to  the net mass inflow rate at small radii 
being one order of magnitude
lower than the inflow rate at large radii.

We conclude that both rotation and position-dependent $T_{\rm R}$
lead to qualitative  and qualitative changes in the flow.
Most prominent of them are weaker collimation and fragmentation of 
the outflow in cases with rotations,
and production of an thermal equatorial outflow in cases
with position-dependent $T_{\rm R}$.

\subsection{Complex Case}

A realistic model of accretion flows should include many physical effects.
Here, we focused on the role of gas rotation and position-dependent 
$T_{\rm R}$. We consider our simulations just as exploratory tests.
These tests support the notion that AGN radiation can play a very important
role in determining the ionization structure and dynamics
of matter on sub- and parsec-scales.
We finish the presentation of our models
with showing results for two runs that illustrate
how complex the flow dynamics can be even in a very simple
set-up as the one we focus on here.

One of our motivations is to understand  gas dynamics
in broad line regions (BLRs) and Narrow Line Regions (NRL) so
characteristic to AGN. These regions are thought be made of cold gas clouds 
moving randomly or nearly randomly and having a small filling factor
(e.g., Krolik 1999 and reference therein). 
Formation, evolution and other key aspects
of these clouds are not well understood. Our simulations
show that an accretion flow which is initially smooth and spherical
can break into inflows and outflows. In cases with rotation,
we have seen outflows fragmented due to line force
and X-ray irradiation. These results raise the following
question: can we produce many cold clouds with a small filling factor.
The answer seems to be yes as we show in Figs. 12 and 13, and 14.

Figs. 12, 13, and 14 compare results for case C with 
$\Gamma=0.9$ and $\rho_o=10^{-20}$~g~cm$^{-3}$.
Left panels in Fig. 12 show results for run Crgd.
The density and temperature distribution and as well as some other properties of this run
differ a lot from other runs shown here. The main difference  is a much larger
dynamical range in the temperature and density plus the fact that the flow
is far from reaching any steady state. However, close inspection of the results
from run Crgd show that the flow pattern is similar to
that seen in most other runs: there is an equatorial inflow and
a bipolar outflow. The latter is not fragmented because it is radial
in agreement with
our explanation of the outflow fragmentation.
The large dynamical range in the density and temperature in run Crgd is 
a simple
consequence of the higher density at the outer boundary. A high density
gas can cold much faster than a low density gas. In this run the gas
is also heated by shocks because for $\Gamma=0.9$ radiation pressure 
of electrons
alone can drive a powerful and broad outflow that collides with 
an inflowing gas.

As we discussed in \S2.2, an accretion flow at large radii does not have to be
heated by the central radiation source. The lower left panel in Fig.~12
illustrates this point as one can see that near the equator a good fraction
of the inflow is relatively cold. One of the prediction
of such a model is that for a wide range of the inclination  angle,
this cold gas should produce absorption lines
redshifted with respect to the systematic velocity. However,
such lines are not being observed. 

The problem of large scale accretion of cold gas can be easly overcome
by introduction of non-central source of heating. 
Right panels in Fig.~12, show results for 
run Crbgd which is a rerun of model Crgd with the X-ray background radiation
(see \S2.2).

In run Crbgd, there are no large regions of shock heated gas. The only
region where shock heating is important is in a narrow polar region
of a low density. The background radiation heating in this run keeps the gas 
from rapid cooling which in turn can lead to abrupt turn-on of 
line driving and strong expulsion of gas as seen in run Crgd. 

Comparing runs Cr and Crbgd, we see that an outflow tends to fragment more
if the gas density is higher. The density and temperature maps show 
that the outflow
occupies a relatively large fraction of the computational domain.
However, Fig. 14 shows that this outflow does not change much the overall
mass budget. As in run Cr, this is caused by rotation that reduces
the amount of gas that is pushed toward the polar region where it can be
siphoned off.


\section{Conclusions}

We have calculated a series of models for rotating flows 
that are under the influence of supermassive BH gravity and
radiation from an accretion disk surrounding the BH. 
We seek to determine self-consistently 
what fraction of  the flow is gravitationally captured by the BH
or what fraction is driven away by thermal expansion and radiation pressure.
This work is an extension of the work presented  in
Paper~I, where  nonrotating flows were studied. Here, we 
consider effects of rotation and of a position-dependent radiation 
temperature, density at large radii, and an uniform X-ray background radiation.

As in the non-rotating case, the rotating flow settles 
into a configuration with two components (1) an equatorial inflow and
(2) a bipolar inflow/outflow with the outflow leaving the system
along the pole. However, the rotating flow does not reach
a steady state. In addition, rotation reduces the outflow collimation and 
the outward flux of mass and kinetic energy. Moreover rotation 
increases the outward flux of the thermal energy and
can lead to fragmentation and time-variability of the outflow.
In future, we plan to check whether thermal instability 
can contribute to fragmentation and time-variability of the outflow.
As expected, rotation does not change much the mass inflow rate 
through the outer boundary. 

In our model, the radiation comes from an UV emitting disk and X-ray emitting 
spherical corona. As a result, the radiaton temperature is position-dependent:
in the polar region radiation is dominated by a softer disk component whereas 
near the equator, radiation is dominated by a harder corona component. 
The two main changes due this position-dependence are: 
(i) an increase in the power of the outflow in the polar region and 
(ii) development of a large scale thermally driven outflow in 
the equatorial region.

Overall, we conclude that our exploratory study provides an additional 
support to the idea  that AGN radiation can significantly change
gas dynamics and photoionzation structute on sub-parsec- and parsec-scales.
As it have been discussed and shown in the past, we found that AGN radiation
can significantly reduce the rate at which 
the central BH is fed with matter (e.g., Figs. 8 and 13).
This result should not depend on the inner radius of the computional domain
because most of the outflow is launched from a radius
larger than the inner radius of the computational domain.
However, we note when reducing the inner radius of the computional domain
one should also consider additional processes, in particular, disk accretion
and disk winds and jets. Thus our mass inflow rate should
be viewed only as an upper limit for the BH accretion rate.

Our results emphasize also a little appreciated feature, 
i.e., AGN radiation can drive a non-spherical, multi-temperature and 
very dynamic flow pattern. This result may have implications for 
the problem of AGN feedback and the problem of the origin, geometry and physics
of NLR and BLR. 

\acknowledgements

This work is supported by NASA through grants 
HST-AR-10680 and HST-AR-11276
from the Space Telescope Science Institute, 
which is operated by the Association of Universities for Research 
in Astronomy, Inc., under NASA contract NAS5-26555.

\clearpage

\begin{table*}
\tiny
\begin{center}
\caption{ Summary of results}
\begin{tabular}{l c c c c c c c c c c c c c c c c } \\ \hline \hline
   &     & & & & & &   &  &      &   &   &     &                  &   &    \\
Run & $\Gamma$ & $f_{\rm disk}$ & $f_\ast$ & $f_{\rm UV}$ & $f_{\rm X} $ & $T_{\rm R}$ & ${\cal F}_{\rm X,b}$ &$T_{\rm 0}$  & $\rho_{\rm 0}$ & $r'_{\rm c}$ &$\MDOT_{\rm in}(r_{\rm o})$ & $\MDOT_{\rm net}(r_{\rm i})$  & $\MDOT_{\rm out}(r_{\rm o})$   &     $v_r$         & $P_{\rm k}({\rm r}_o)$        &  $P_{\rm th}({\rm r}_o)$        \\ 

   &      &     &      &    &     &(1)      &(2)&(1)  &(3)&(4)&(5)   &(5)     &(5) &(6)           &(7) &(7)   \\ \hline

A  & 0.6  &  0.5& 0.5  & 0.5& 0.5 &4        &0  &1    & 1 & 0 & -4   & -1    & 3  & 700           & 2   & 4        \\
Ax & 0.6  &  0.5& 0.5  & 0.5& 0.5 &4.8-14.5 &0  &1    & 1 & 0 & -0.8 & -0.1  & 0.7& 400           & 0.1 & 2        \\

B  & 0.6  &  0.8& 0.2  & 0.8& 0.2 &4        &0  &1    & 1 & 0 & -8   &-3     & 5  & 4000          & 100 & 0.8        \\
Br & 0.6  &  0.8& 0.2  & 0.8& 0.2 &4        &0  &1    & 1 &300& -8   &-5     & 3  & 1300          & 4   & 0.5        \\

B1 & 0.6  &  0.8& 0.2  & 0.8& 0.2 &4        &0  &1/10 & 1 & 0 & -0.5 &-0.09  &0.41& 1500          & 0.5 & 0.8        \\
B2 & 0.6  &  0.8& 0.2  & 0.8& 0.2 &4        &0  &1/3  & 1 & 0 & -2   &-0.4   &1.6 & 1700          & 2   & 2        \\
B3 & 0.6  &  0.8& 0.2  & 0.8& 0.2 &4        &0  &3    & 1 & 0 & -9   &-5     & 4  & 400           & 3   & 0.8        \\

C  & 0.6  &  0.95&0.05& 0.95&0.05 &4        &0  &1    & 1 & 0 & -9   & -1     & 8  & 6700          & 700 & 0.03    \\ 
CR & 0.6  &  0.95&0.05& 0.95&0.05 &4        &0  &1    & 1 &300& -10  & -8     & 3  & 600          & 3  & 0.2    \\ 
Cr & 0.6  &  0.95&0.05& 0.95&0.05 &4        &0  &1    & 1 &300& -10  & -4     & 6  & 1000          & 10  & 0.2    \\ 
Cx & 0.6  &  0.95&0.05& 0.95&0.05 &0.3-14.5 &0  &1    & 1 & 0 & -11  & -0.15  & 10.85&7000         & 300 & 0.03    \\ 
Crx& 0.6  &  0.95&0.05& 0.95&0.05 &0.3-14.5 &0  &1    & 1 &300& -11  & -2     & 3  & 500           & 10  & 0.01    \\ 

Crgd & 0.9  &  0.95&0.05& 0.95&0.05 &4        &0  &10    & 10 &300& -115 & -5     & 110& 2500          &1000 & 10.    \\ 
Crbgd& 0.9  &  0.95&0.05& 0.95&0.05 &4        &1.2&10    & 10 &300& -150 & -120   & 30 & 3700          & 30  & 5.    \\ 
\hline
\end{tabular}
\tablecomments{(1) in $2\times10^7$~K,
(2) in $10^{7}$~erg~cm$^{-2}$~s$^{-1}$,
(3) in $10^{-21}$~g~cm$^{-3}$,
(4) in $r_\ast$,
(5) in $10^{25}$~g~s$^{-1}$,   
(6) in $\rm km~s^{-1}$,
(7) in $10^{40}$~erg~s$^{-1}$.
We use the following convention to label our runs:
the first character refers to the values of $f_{\rm X}$ and $f_{\rm UV}$:
A is for $f_{\rm UV}=0.5$  and $f_{\rm X}=0.5$, 
B is for $f_{\rm UV}=0.8$  and $f_{\rm X}=0.2$, and
C is for $f_{\rm UV}=0.95$  and $f_{\rm X}=0.05$. Runs A, B, and C,
are the fiducial runs.
If the first character
is followed by a lower case letter or letters or a number, 
it means that it is the same
run, but modified by introduction of:
rotation (letter 'R' and 'r' if we use eq. 4 or 5, respectively), the position-dependent $T_{\rm R}$ (letter 'x'),
the X-ray background radiation (letter 'b'), a higher $\Gamma$
(letter 'g'),
and a  higher $\rho_{\rm 0}$ (letter 'd'). The numbers 1, 2, and 3
correspond respectively to $T_{\rm R}=1/10, 1/3, 3$ in units of
$2\times10^7$~K.}
\end{center}
\normalsize
\end{table*}

\newpage

\begin{figure}
  \begin{tabular}{ccc}
    \includegraphics[clip,width=0.30\textwidth,angle=90]{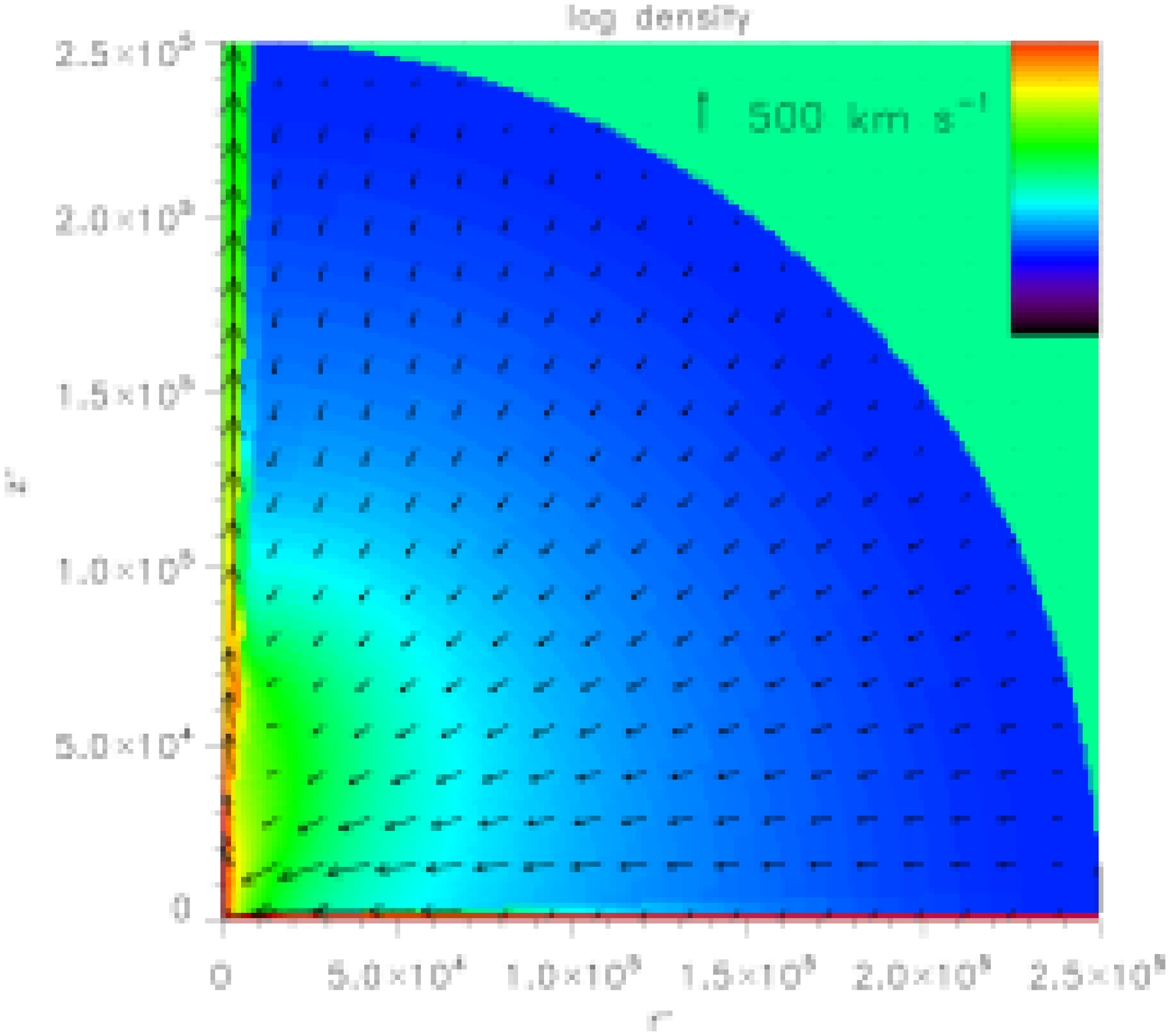}&
    \includegraphics[clip,width=0.30\textwidth,angle=90]{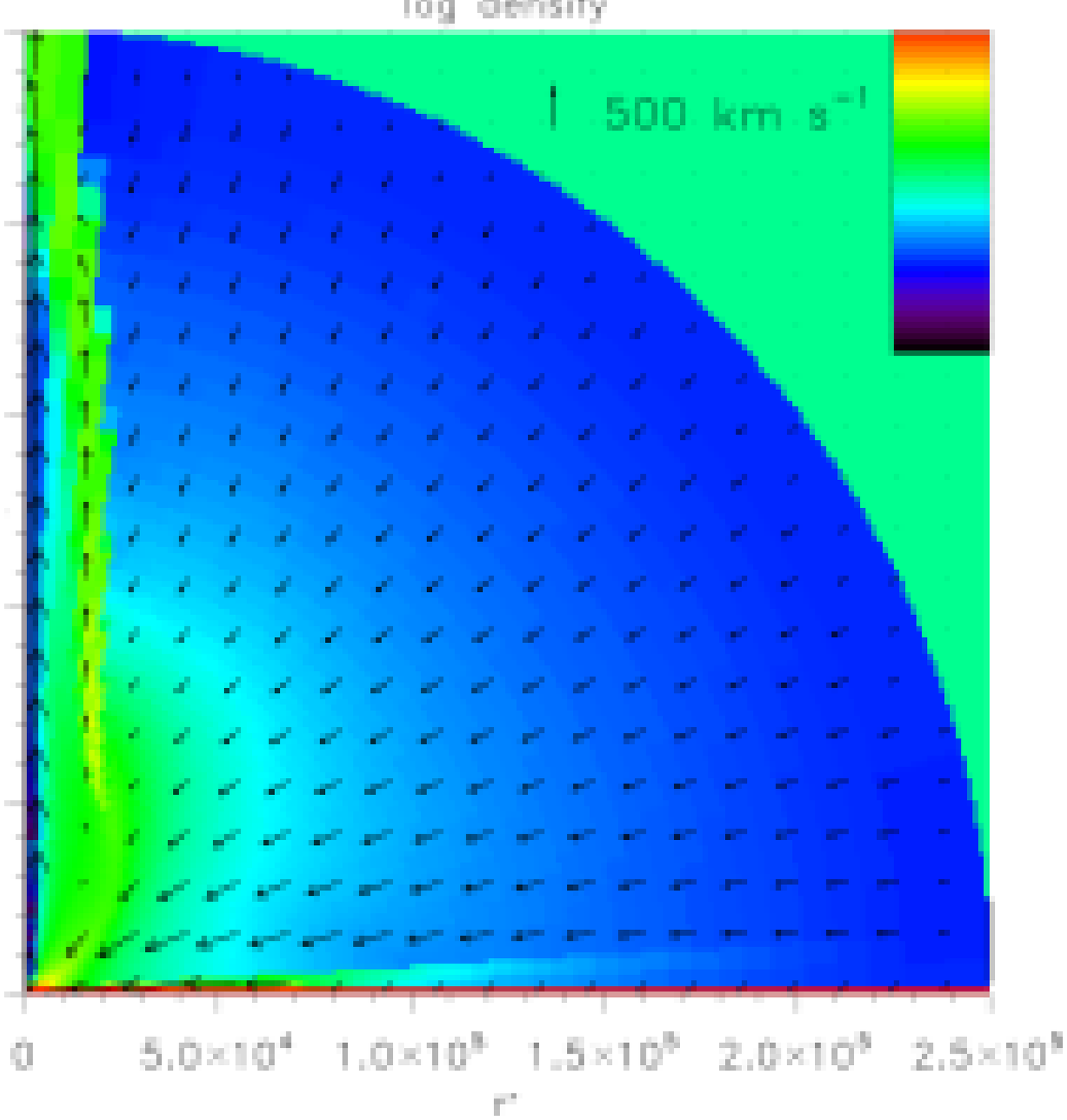}&
    \includegraphics[clip,width=0.30\textwidth,angle=90]{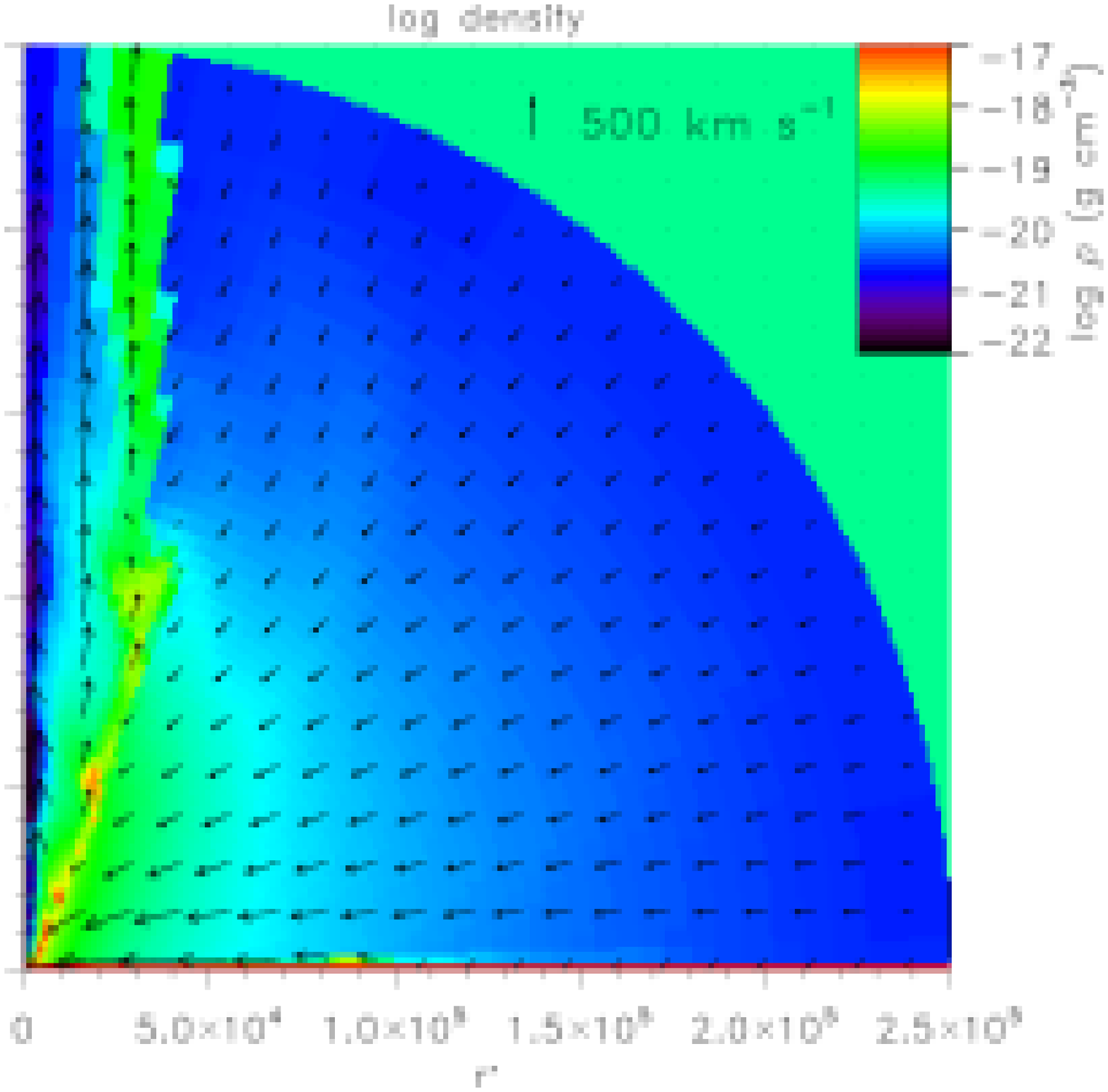}\tabularnewline
    \includegraphics[clip,width=0.30\textwidth,angle=90]{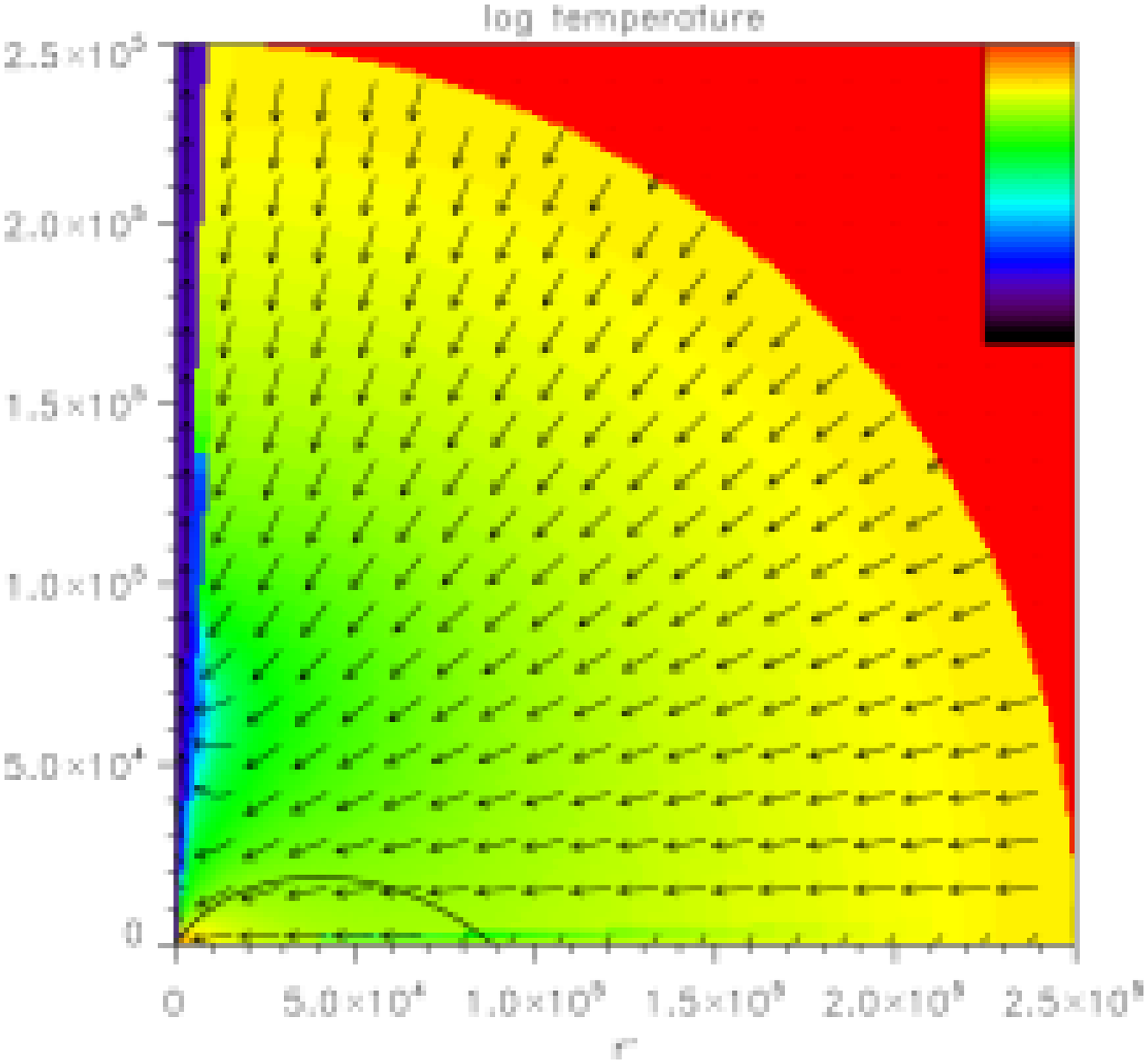}&
    \includegraphics[clip,width=0.30\textwidth,angle=90]{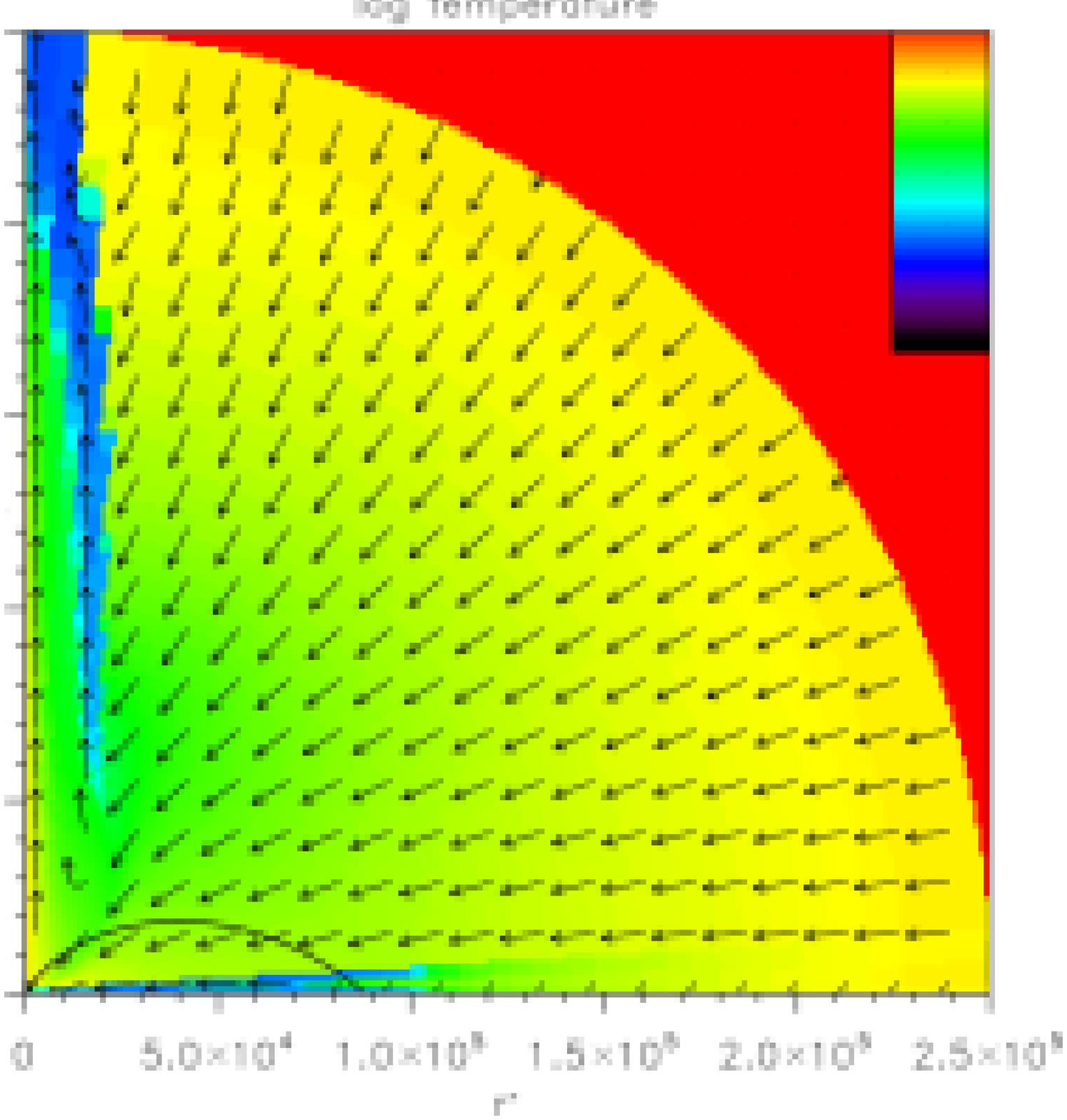}&
    \includegraphics[clip,width=0.30\textwidth,angle=90]{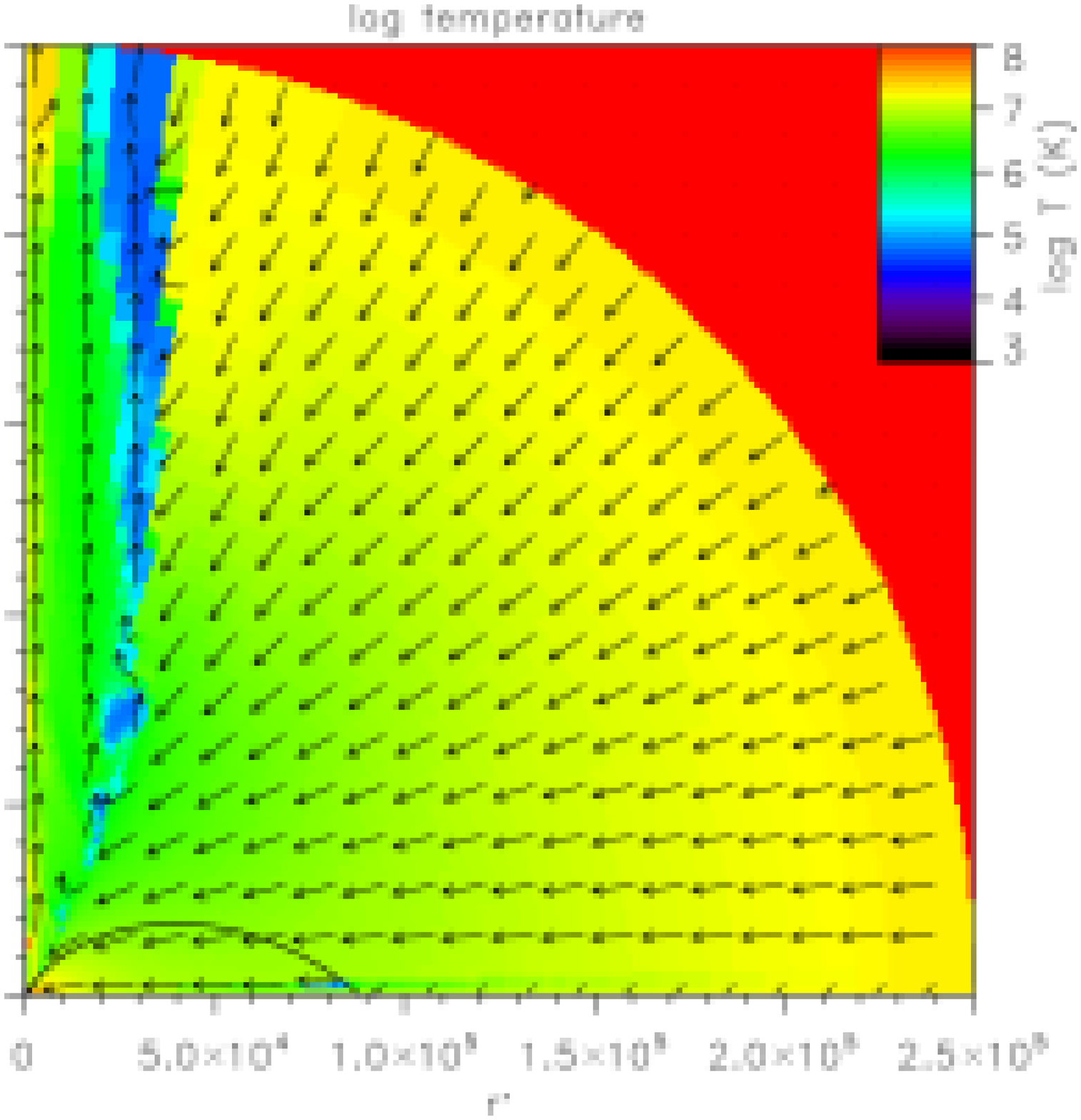}\tabularnewline    
  \end{tabular}
  \vspace{1cm}

%
%
%

\caption{Comparison of the results for runs C, CR, and Cr 
(left, middle, and right column, respectively)
{\it Top row:}
Maps of logarithmic density overplotted by the 
poloidal velocity. For clarity, the arrows
are plotted with
the maximum velocity  set to 1000~${\rm km~s^{-1}}$. The solid curves show are the contours of the Mach number equal one.
{\it Bottom row:}
Maps of logarithmic temperature overplotted by the direction of the 
poloidal velocity. The solid curve in the bottom left corner
marks the position of the Compton radius corrected for the effects
of radiation pressure due to electron scattering (see eq. 19 in Paper~I).
The length scale is in units of the inner disk radius 
(i.e., $r' = r/r_\ast$ and $z' = z/r_\ast$).
}

\end{figure}

\begin{figure}
  \includegraphics[width=0.80\textwidth,angle=90]{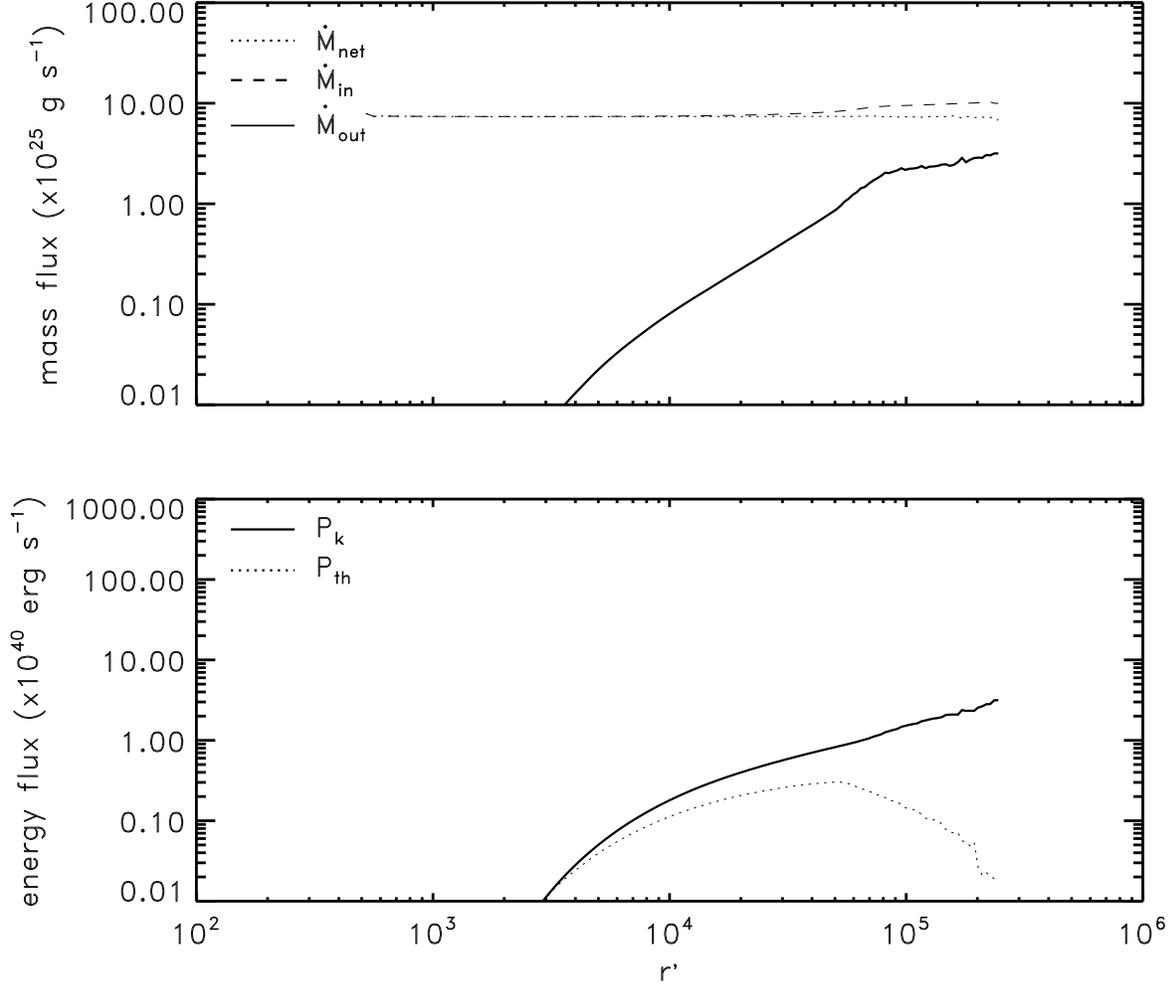}
  \vspace{1cm}
%
\caption{{\it Top panel:} The mass flux rates 
as a function of radius for run CR.
The solid, dashed, and dotted line corresponds to the outflow, inflow, 
and net rates, respectively (see eqs. 20, 21, and 22 in Paper~1 for the formal
definitions). Note that the absolution value 
of the inflow and net rates are plotted because these quantities are negative.
{\it Bottom panel:} The energy fluxes carried out by the outflow
as a function of radius in run CR.
The solid and dashed line corresponds to the kinetic and thermal
energy flux, respectively (see eqs. 23 and 24 for the formal definitions).
The length scale is in units of the inner disk radius 
(i.e., $r' = r/r_\ast$).}
\end{figure}

\begin{figure}
  \includegraphics[width=0.80\textwidth,angle=90]{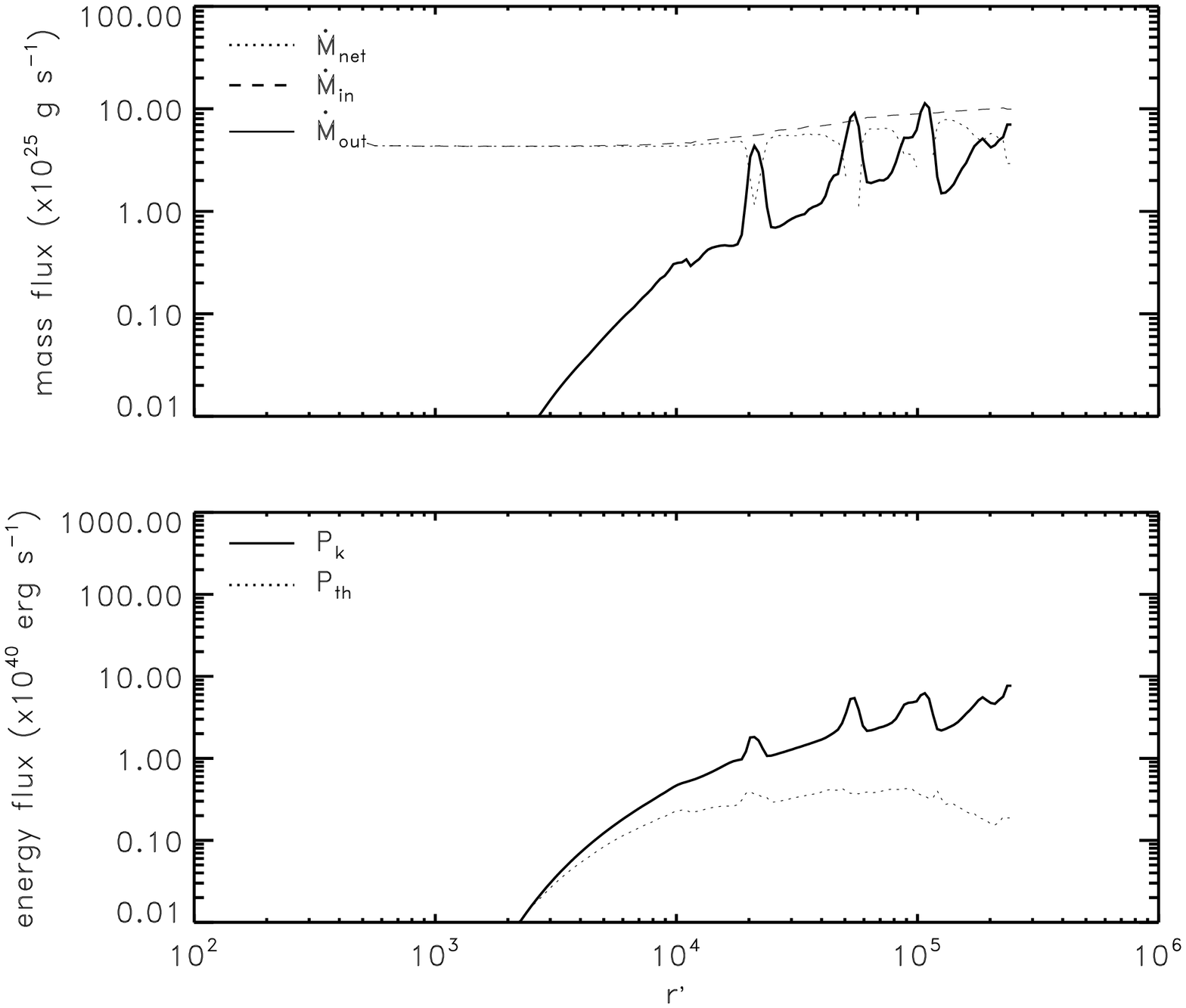}
  \vspace{1cm}

%
\caption{As in Fig.~2, but for run Cr.}
\end{figure}

\begin{figure}
  \begin{tabular}{cccc}
    \includegraphics[clip,width=0.23\textwidth,angle=90]{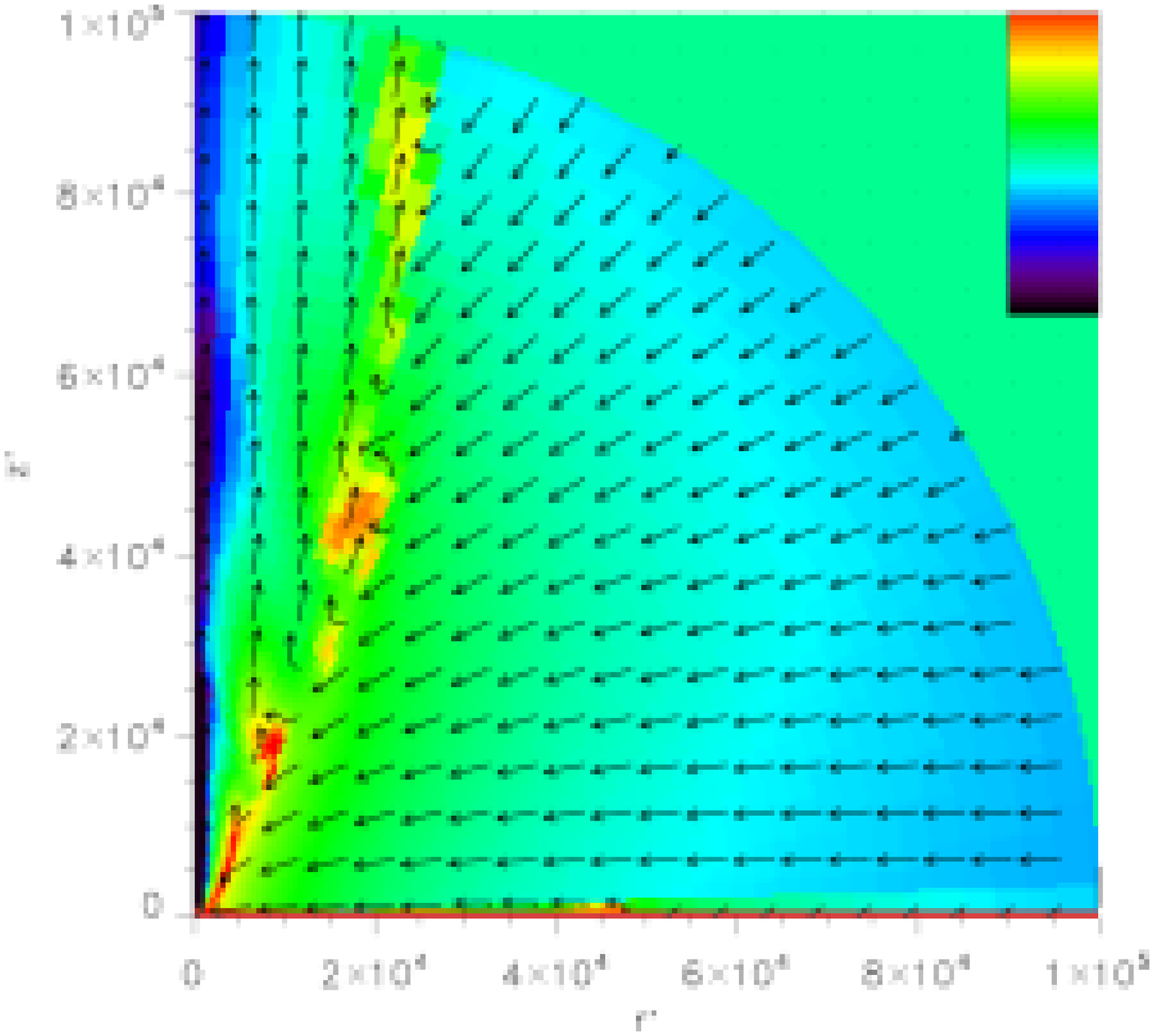}&
    \includegraphics[clip,width=0.23\textwidth,angle=90]{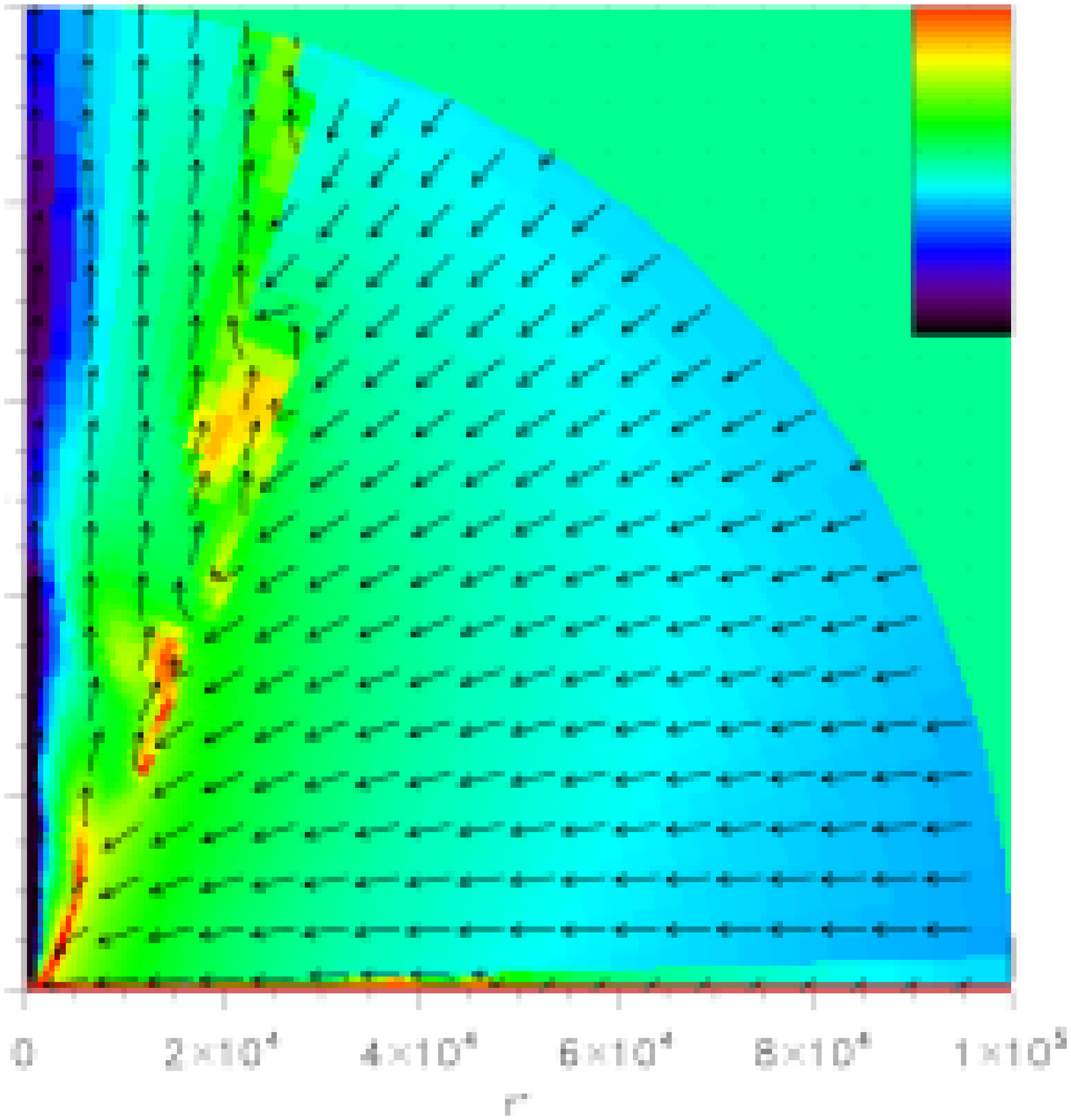}&
    \includegraphics[clip,width=0.23\textwidth,angle=90]{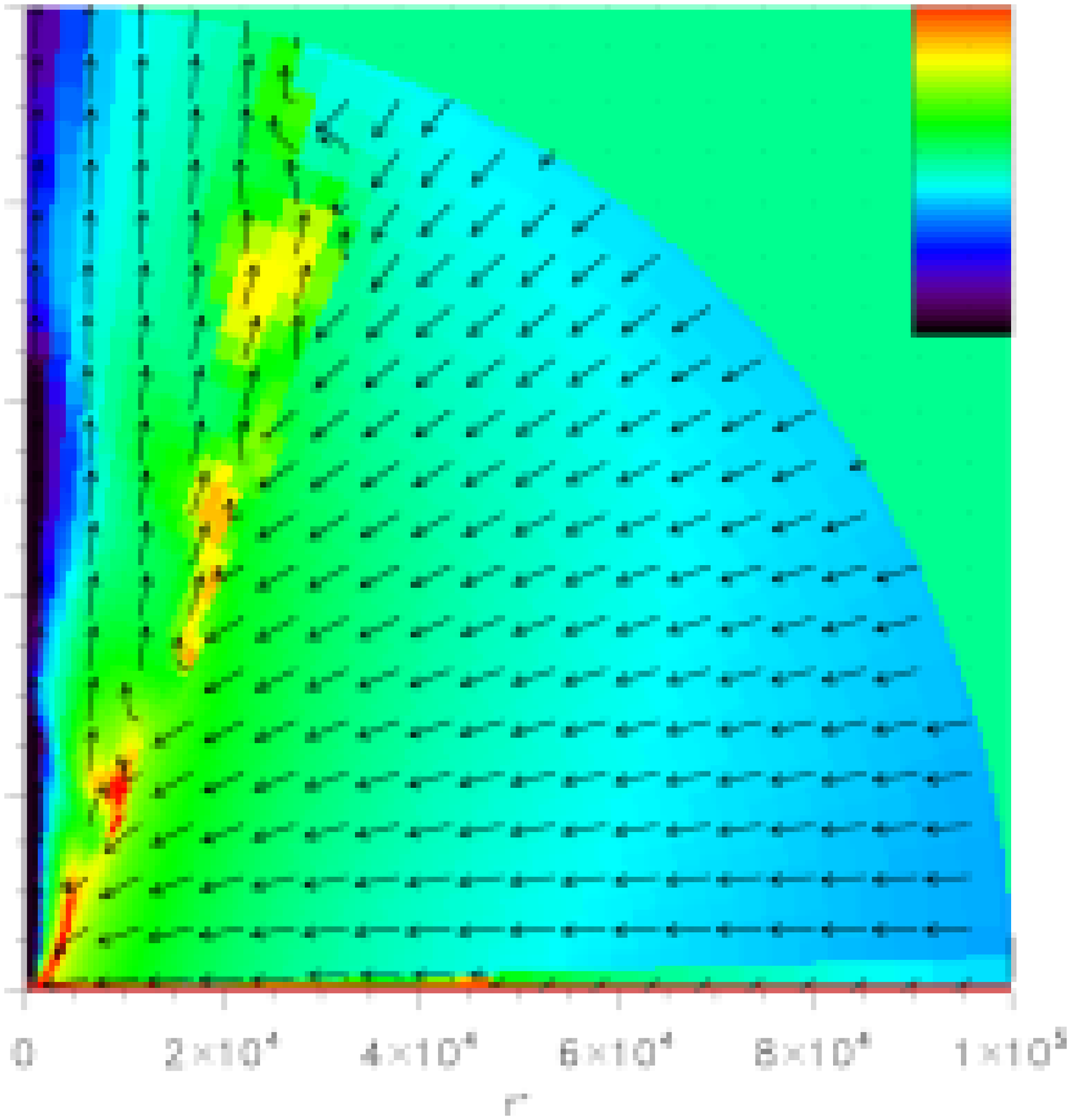}&
    \includegraphics[clip,width=0.23\textwidth,angle=90]{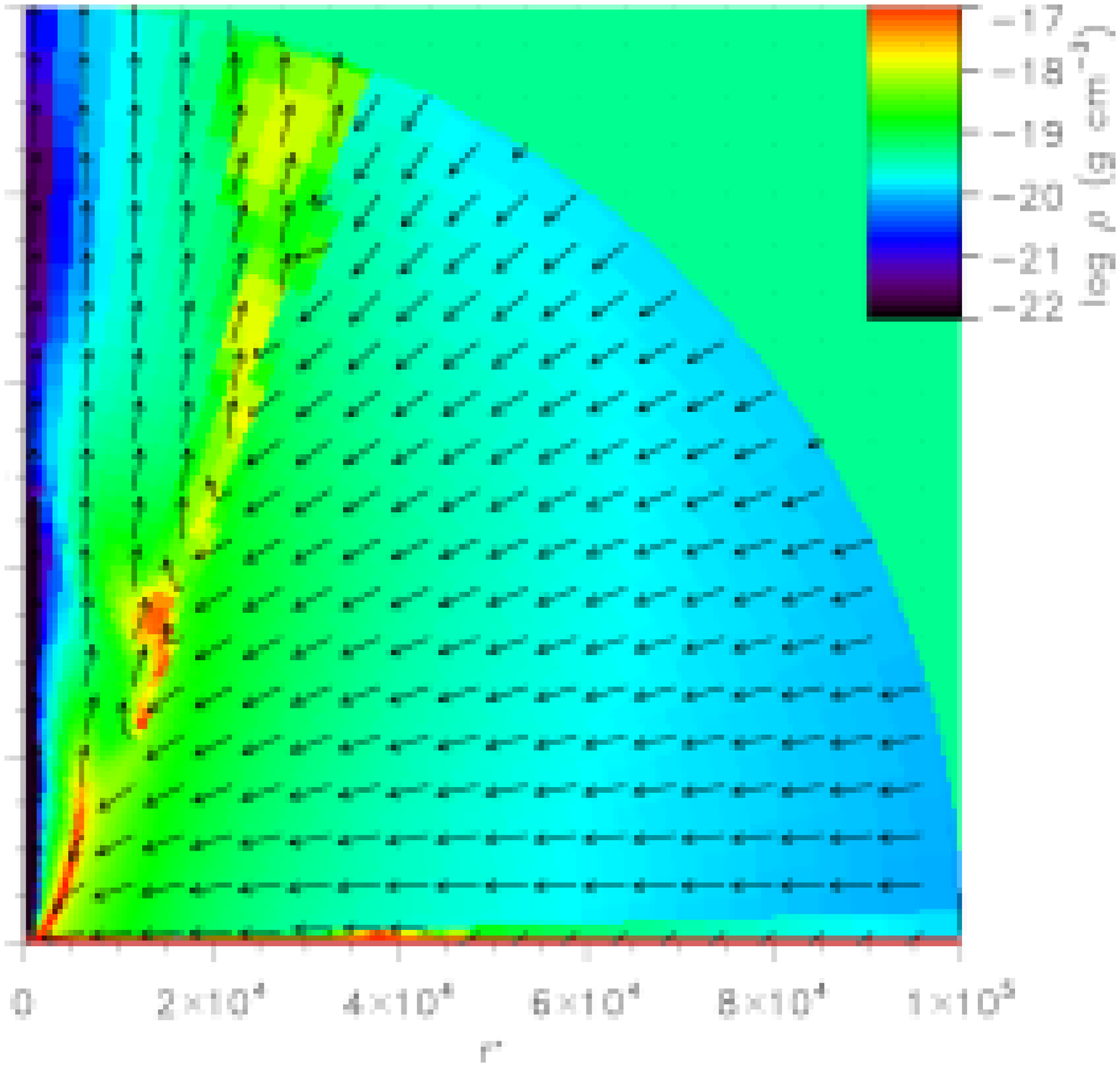}\tabularnewline
  \end{tabular}
  \vspace{1cm}

\caption{A sequence of density maps for the inner part of run CR after
4.95, 5.00, 5.05, and 5.10$\times 10^{12}$~s (from left to right).
Run CR is an example of unsteady flow discussed in detail in section 3.1.
As in fig. 1, the length scale is in units of the inner disk radius 
(i.e., $r' = r/r_\ast$ and $z' = z/r_\ast$). However, the r' and z'
ranges are 2.5 time smaller compared to fig. 1.
}
\end{figure}

\begin{figure}
  \includegraphics[clip,width=0.90\textwidth,angle=0]{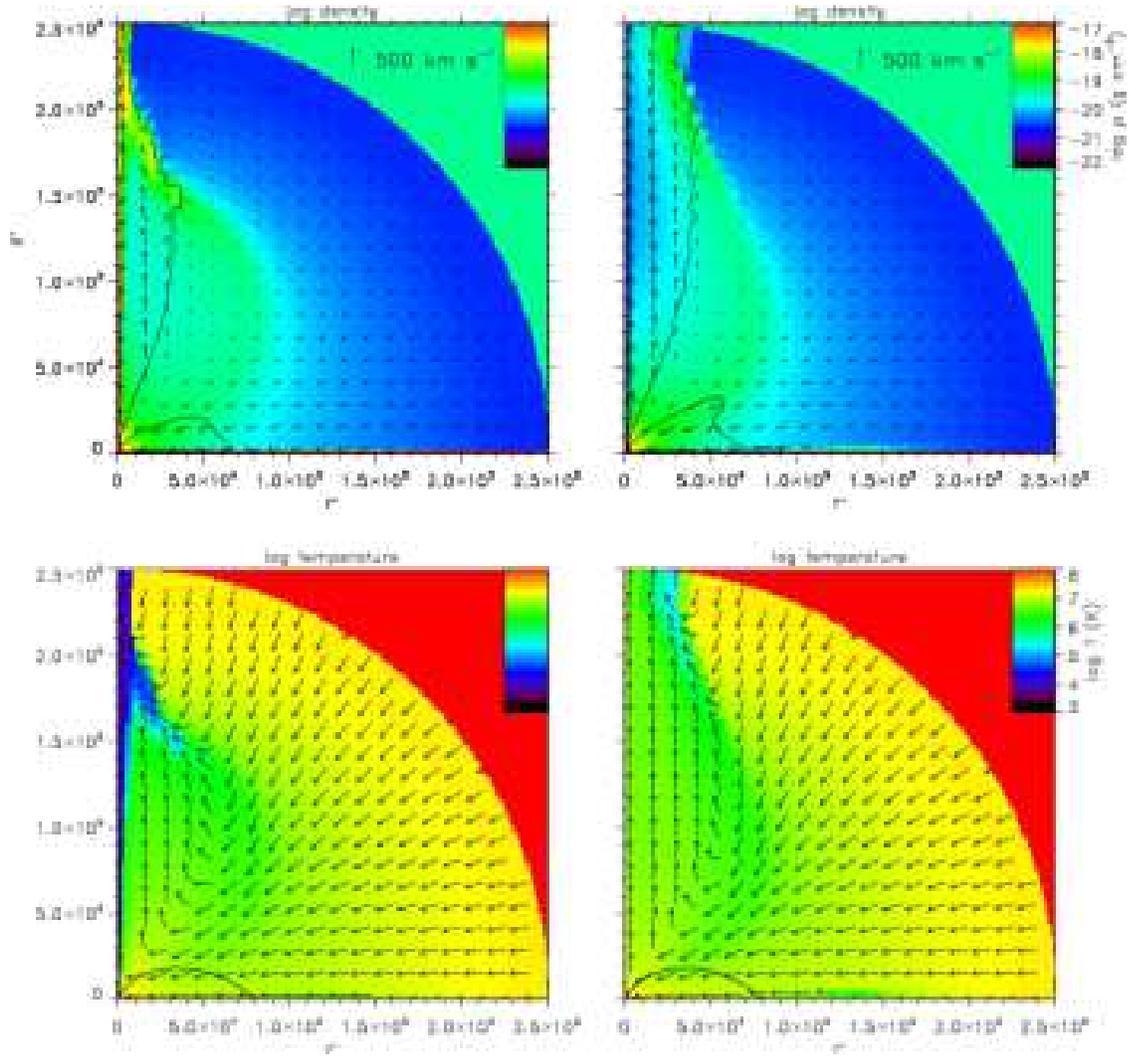}
  \vspace{1cm}

\caption{As in Fig.~1, but for runs B and Br.
}
\end{figure}

\begin{figure}

  \includegraphics[width=0.80\textwidth,angle=90]{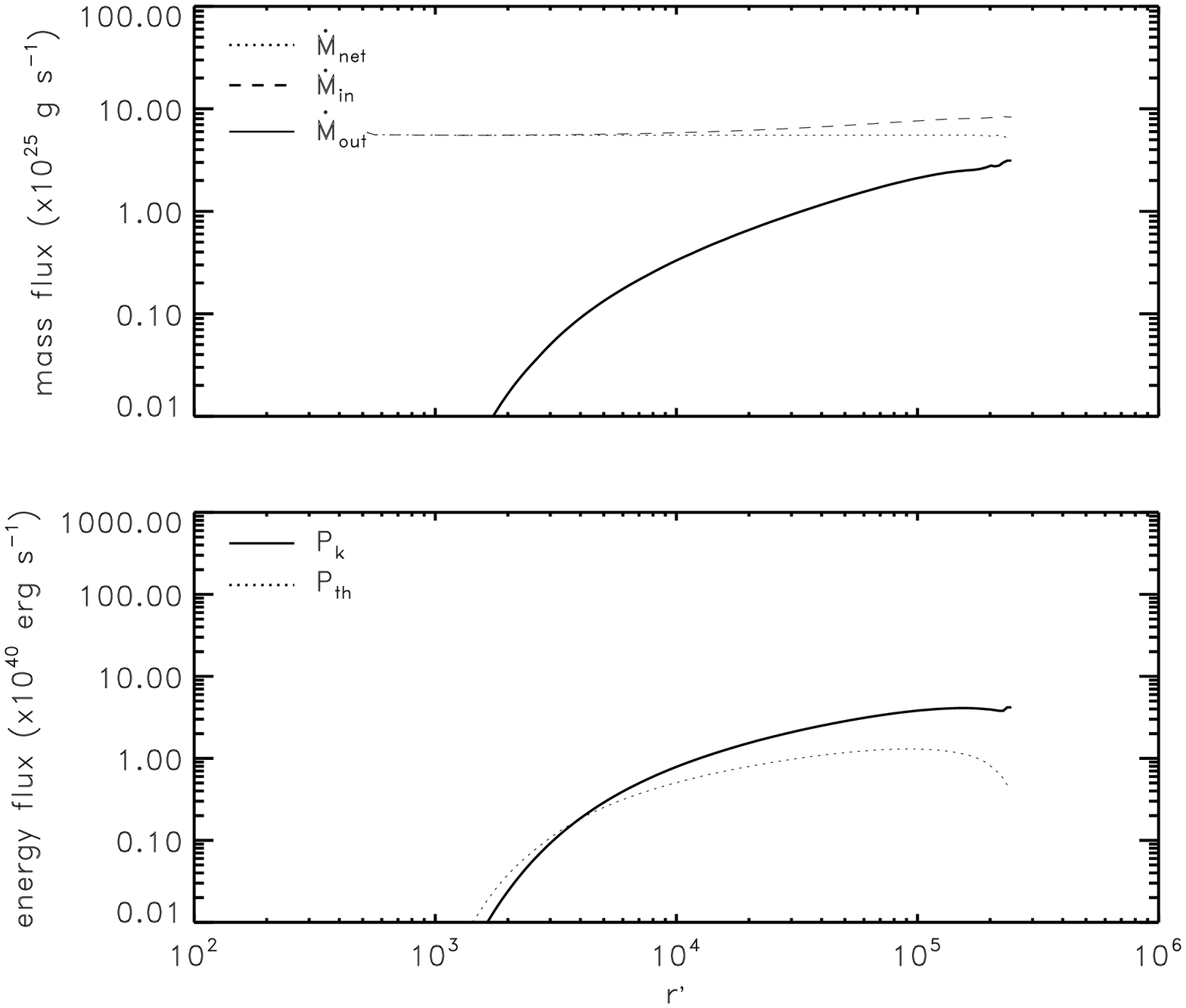}
  \vspace{1cm}
%
\caption{As in Fig.~2, but for run Br.}
\end{figure}

\begin{figure}
  \includegraphics[width=0.90\textwidth,angle=0]{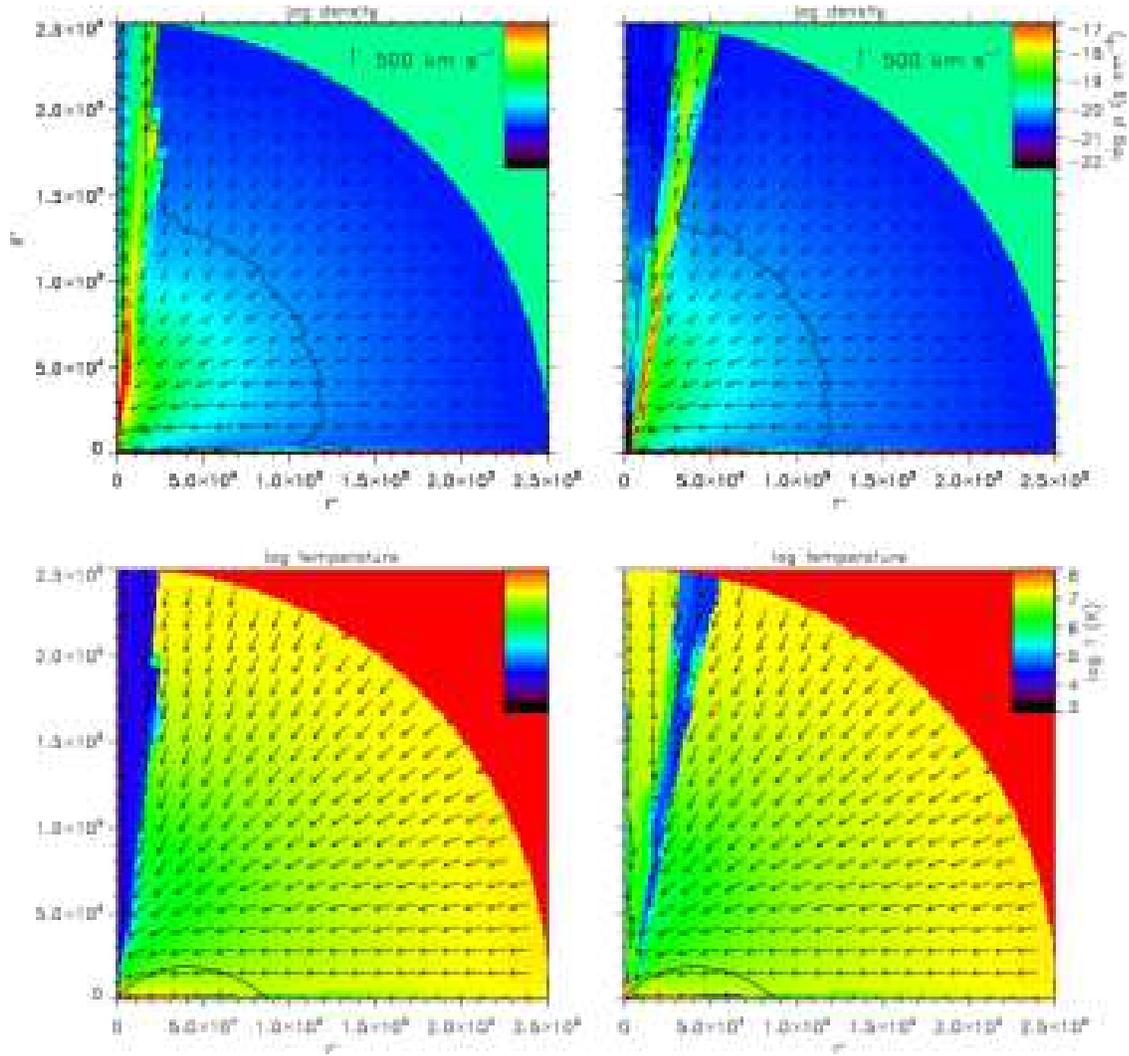}
  \vspace{1cm}

\caption{As in Fig.~1, but for runs Cx and Crx.
}
\end{figure}

\begin{figure}
  \includegraphics[width=0.80\textwidth,angle=90]{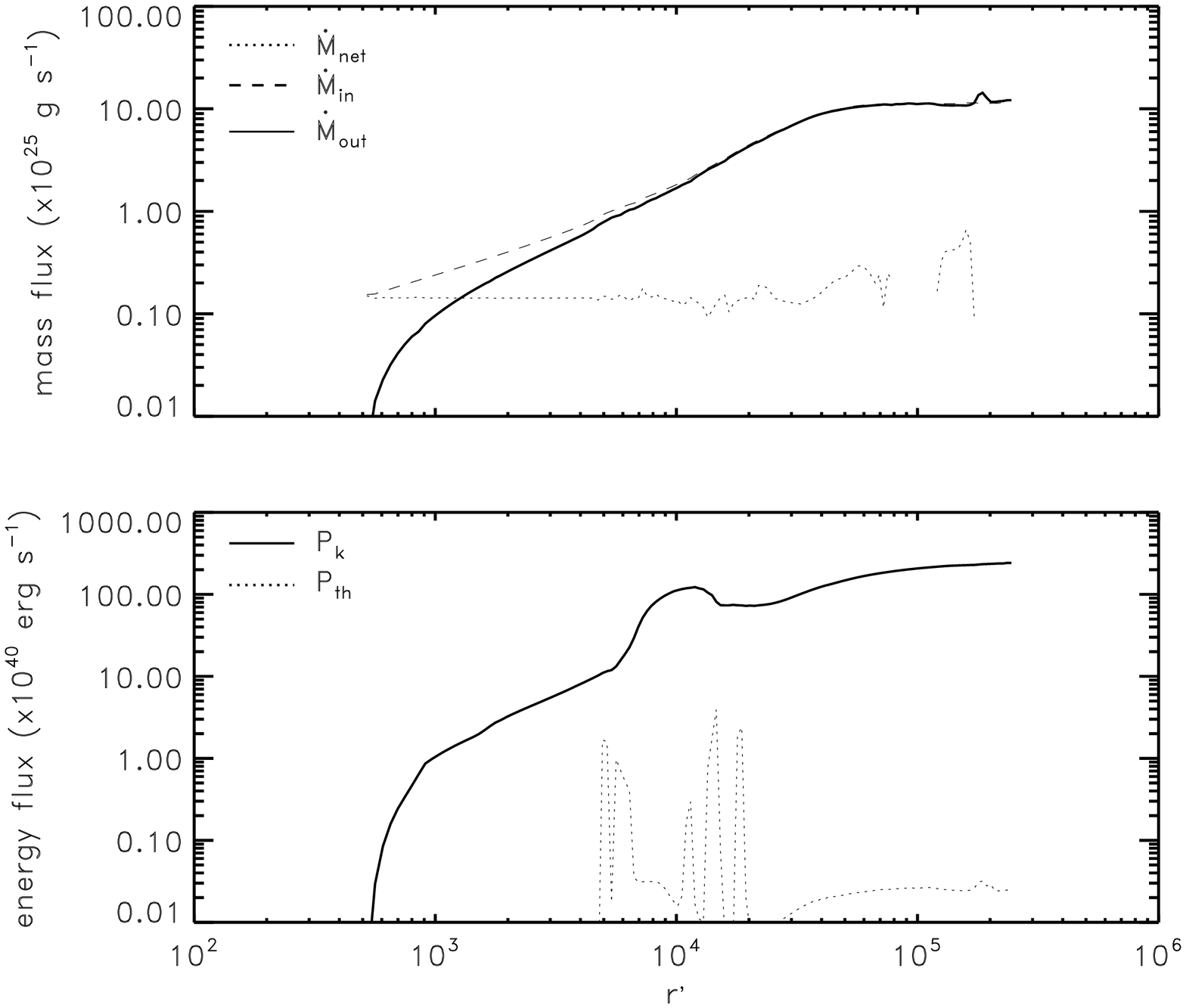}
  \vspace{1cm}
%
\caption{As in Fig.~2, but for run Cx.}
\end{figure}

\begin{figure}
  \includegraphics[width=0.80\textwidth,angle=90]{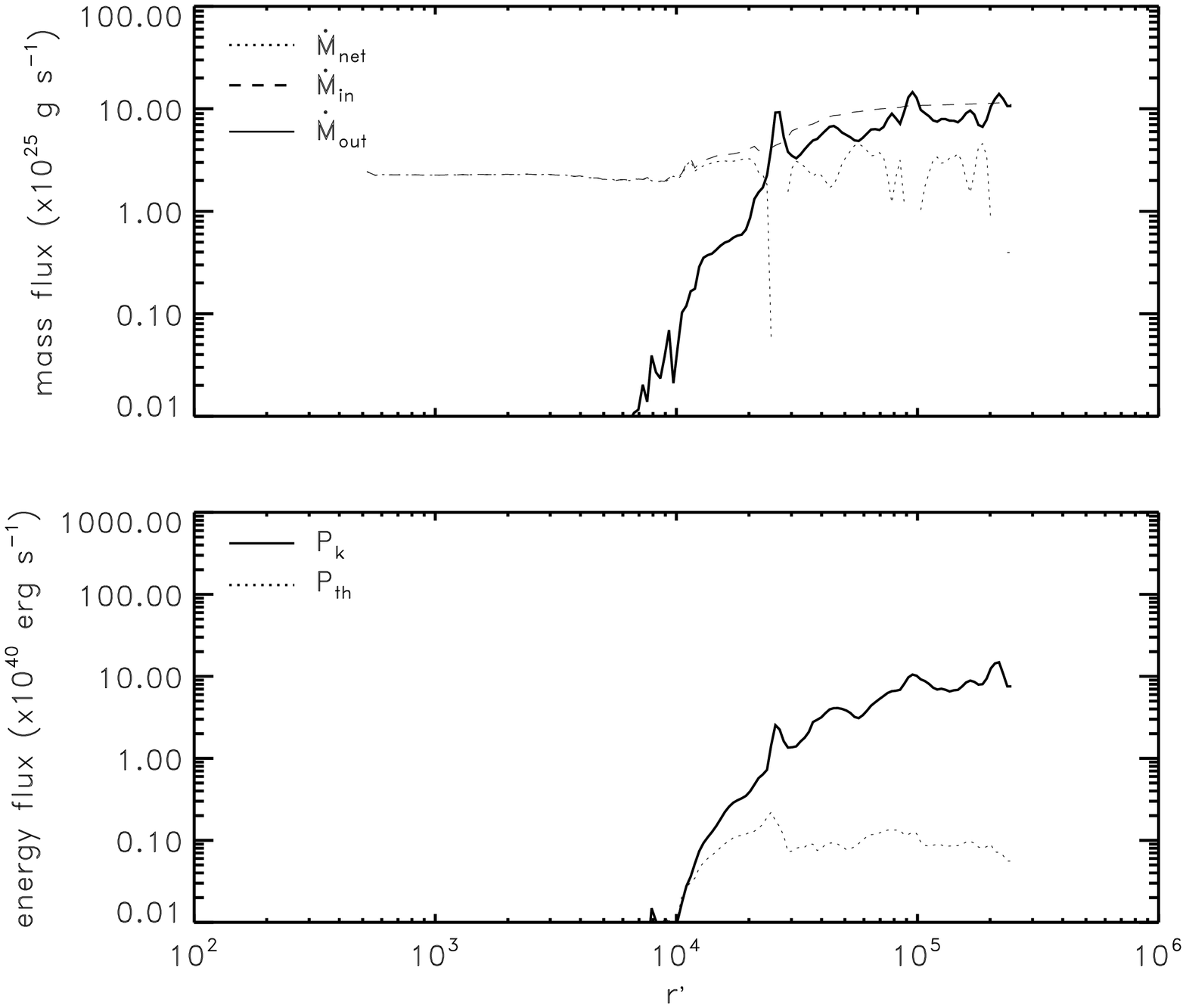}
  \vspace{1cm}

%
\caption{As in Fig.~2, but for run Crx.}
\end{figure}

\begin{figure}
  \includegraphics[width=0.90\textwidth,angle=0]{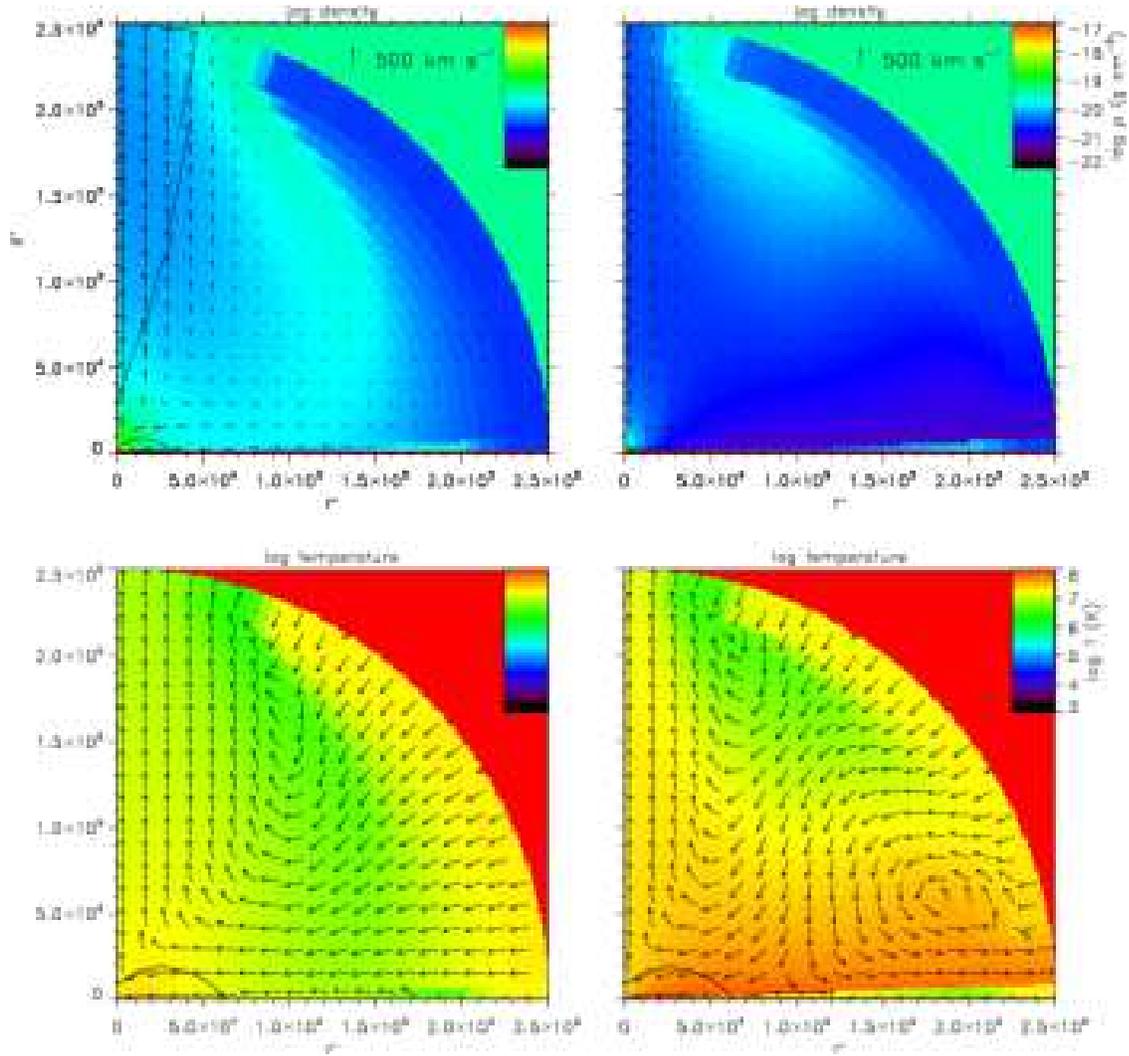}
  \vspace{1cm}

\caption{
As in Fig.~1, but for runs A and Ax.
}
\end{figure}

\begin{figure}
  \includegraphics[width=0.80\textwidth,angle=90]{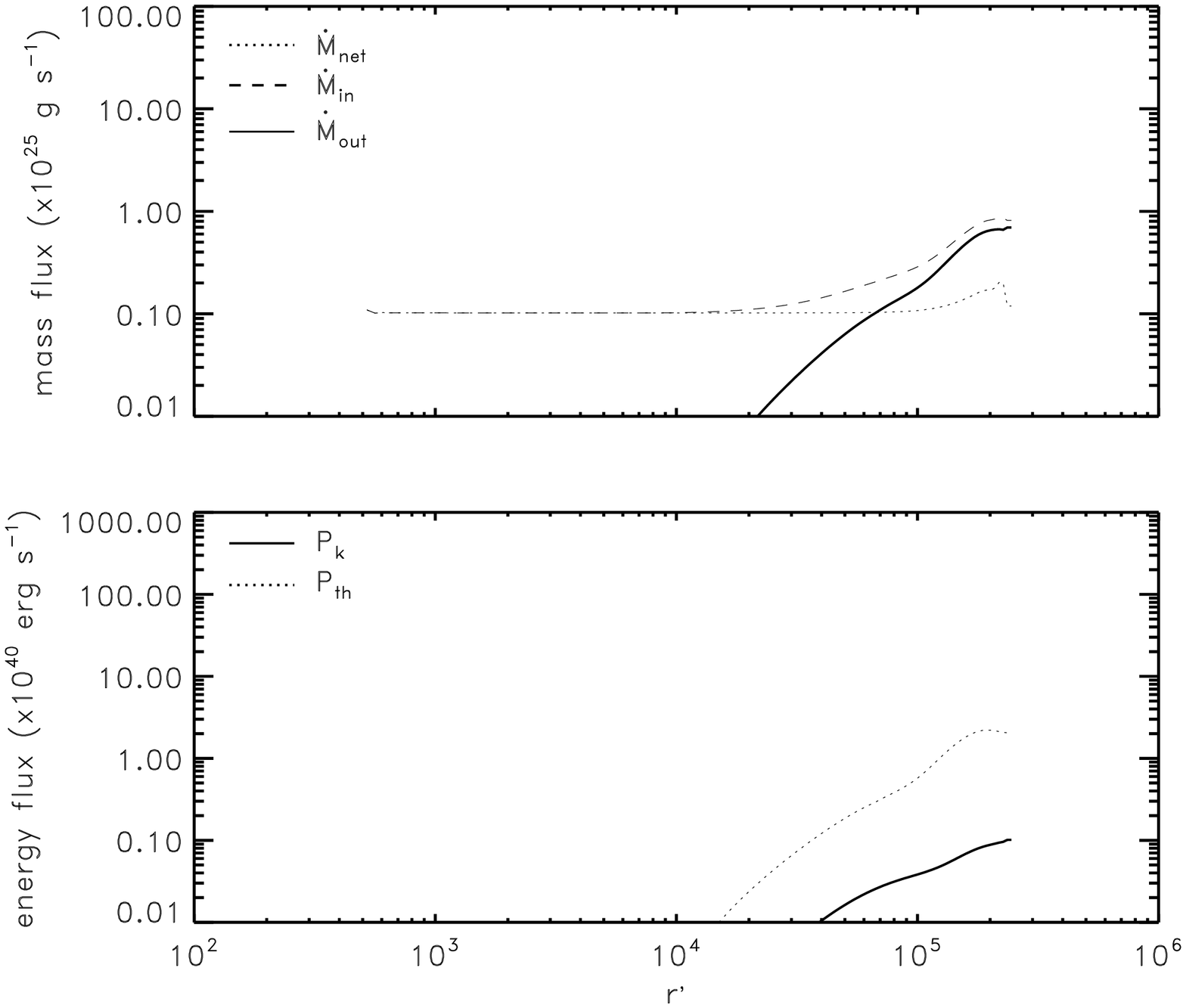}
  \vspace{1cm}

%
\caption{As in Fig.~2, but for run Ax.}
\end{figure}

\begin{figure}
  \includegraphics[width=0.90\textwidth,angle=0]{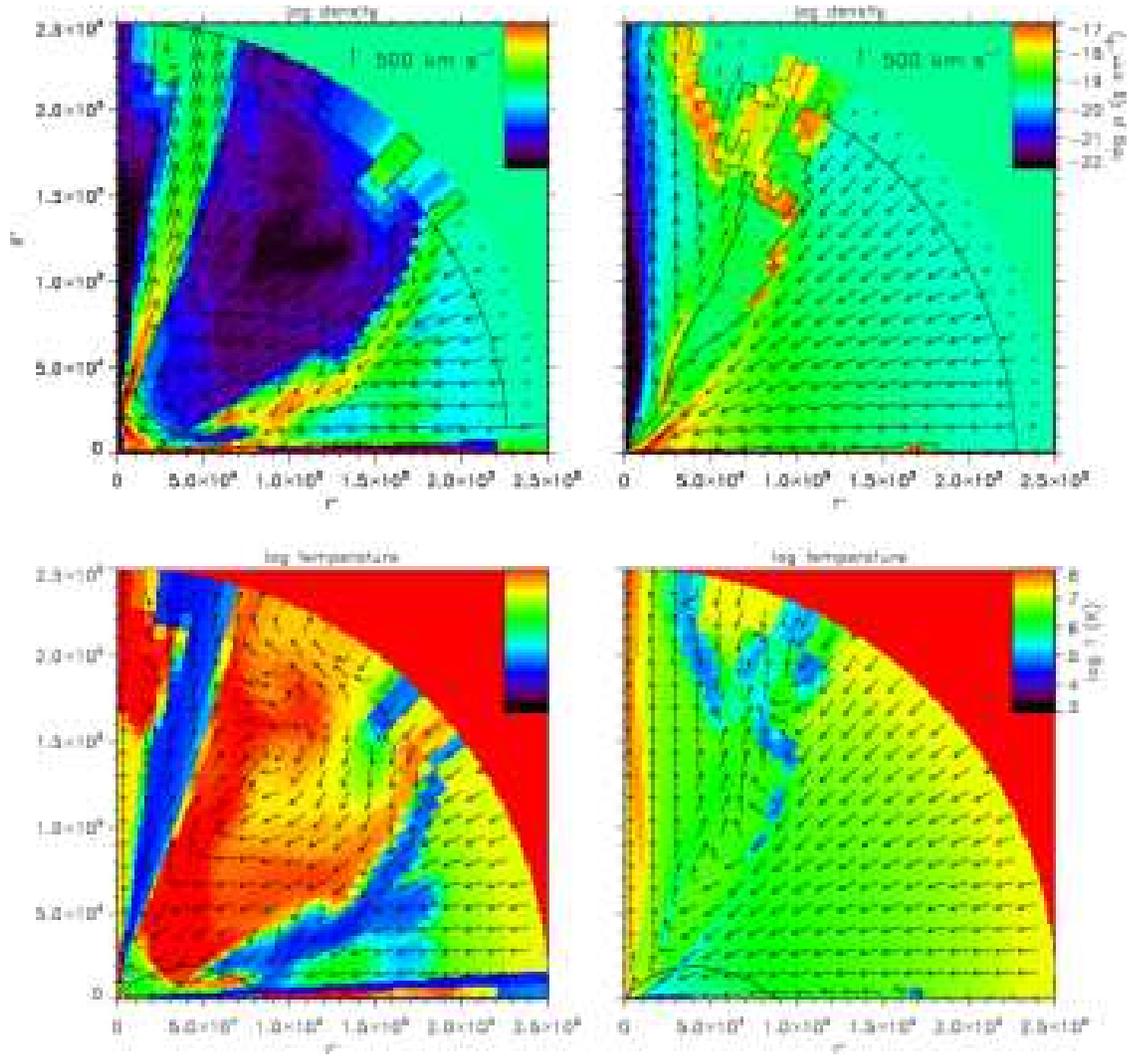}
  \vspace{1cm}

\caption{As in Fig.~1, but for runs Crgd and Crbgd.
}
\end{figure}

\begin{figure}
  \includegraphics[width=0.80\textwidth,angle=90]{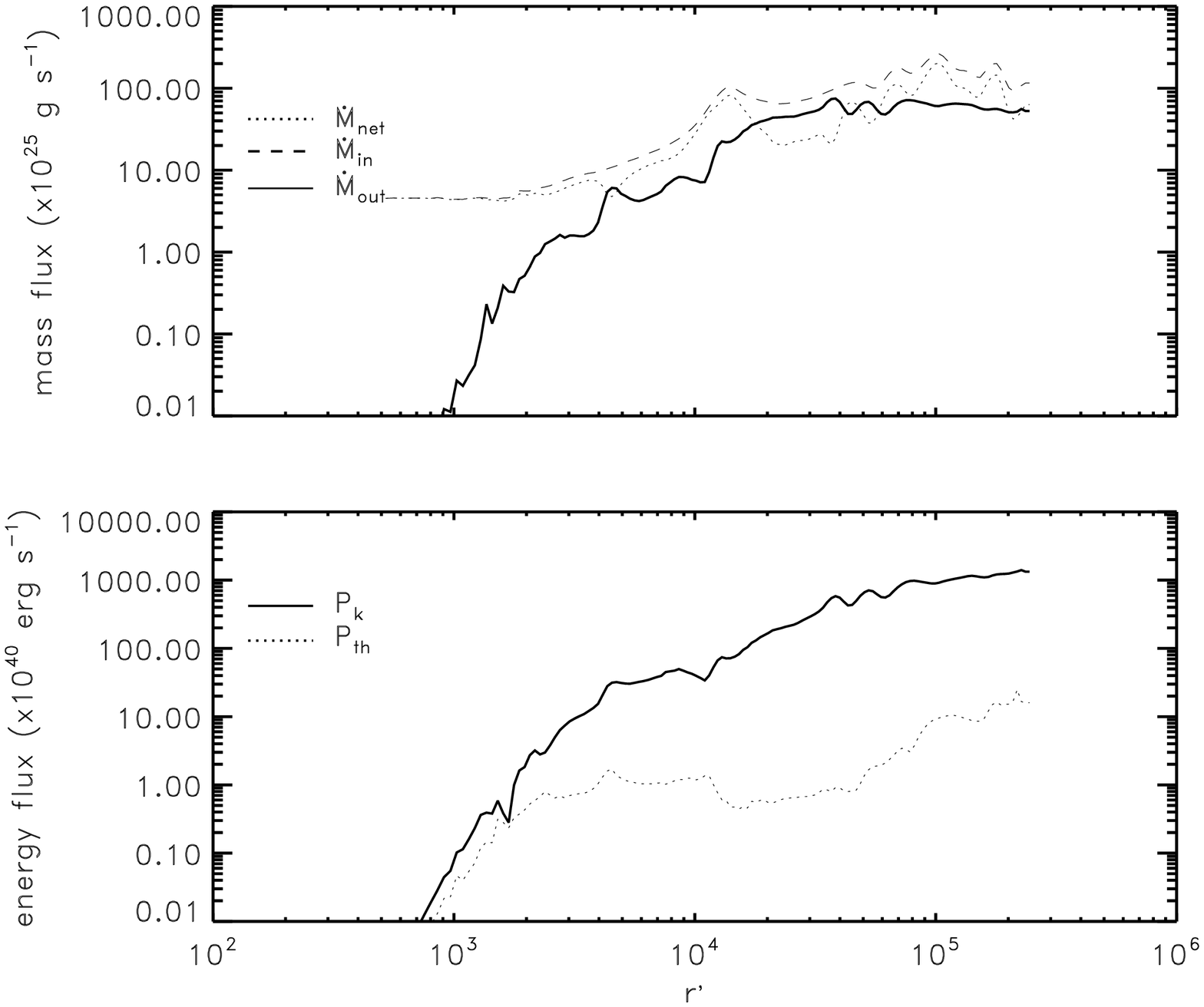}
  \vspace{1cm}

%
\caption{As in Fig.~2, but for run Crgd.
Note an increase range
along the y-axis compared to Fig~2.}
\end{figure}

\begin{figure}
  \includegraphics[width=0.80\textwidth,angle=90]{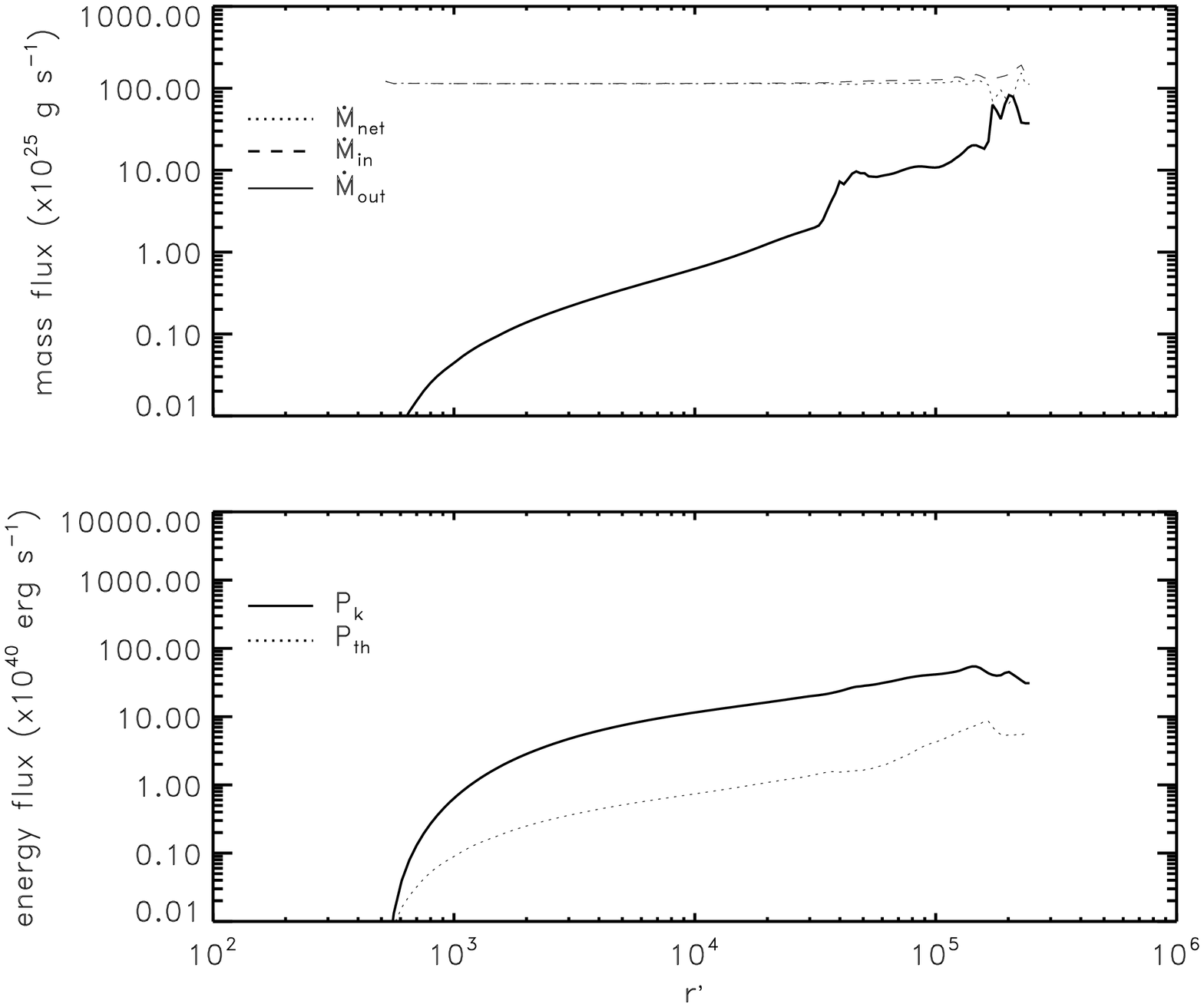}
  \vspace{1cm}

%
\caption{As in Fig.~2, but for run Crbgd.
Note an increase range
along the y-axis compared to Fig~2.}
\end{figure}

\end{document}